\documentclass[aps,pra,reprint,superscriptaddress,floatfix,nofootinbib]{revtex4-2}

\usepackage[english]{babel}
\usepackage[utf8]{inputenc}
\usepackage[T1]{fontenc}

\usepackage{physics}
\usepackage{amsfonts,amsmath,amssymb,accents,braket,mathtools}
\usepackage{dsfont,amsfonts,bm}
\usepackage{array}
\usepackage{enumitem}

\usepackage{placeins}
\usepackage{graphicx}
\graphicspath{{figures/}}
\usepackage{epstopdf}
\usepackage[caption=false, justification=centerlast]{subfig}

\usepackage[colorlinks=true,citecolor=blue,linkcolor=blue,urlcolor=blue]{hyperref}
\usepackage[capitalise]{cleveref}
\usepackage{color,xcolor}
\usepackage{ulem} % underline

\begin{document}

\title{Entangling lattice-trapped bosons with a free impurity: 
	impact on stationary and dynamical properties}

\author{Maxim Pyzh}
	\email{mpyzh@physnet.uni-hamburg.de}
	\affiliation{Center for Optical Quantum Technologies, Department of Physics,
		University of Hamburg, Luruper Chaussee 149, 22761 Hamburg Germany }
\author{Kevin Keiler}
\affiliation{Center for Optical Quantum Technologies, Department of Physics,
	University of Hamburg, Luruper Chaussee 149, 22761 Hamburg Germany }
\author{Simeon I. Mistakidis}
\affiliation{Center for Optical Quantum Technologies, Department of Physics,
	University of Hamburg, Luruper Chaussee 149, 22761 Hamburg Germany }
\author{Peter Schmelcher}
	\email{pschmelc@physnet.uni-hamburg.de}
	\affiliation{Center for Optical Quantum Technologies, Department of Physics,
		University of Hamburg, Luruper Chaussee 149, 22761 Hamburg Germany }
	\affiliation{The Hamburg Centre for Ultrafast Imaging, Universit\"at 
Hamburg, Luruper Chaussee 149, 22761 Hamburg, Germany}

\begin{abstract} 
	
	We address the interplay of few lattice trapped bosons interacting with an impurity atom in a box potential. 
	For the ground state, a classification is performed based on the fidelity allowing to quantify the susceptibility of the composite system to structural changes 
	due to the intercomponent coupling. 
	We analyze the overall response at the many-body level and contrast it 
	to the single-particle level. 
	By inspecting different entropy measures we capture the degree of entanglement and intraspecies correlations for a wide range of intra- and intercomponent interactions and lattice depths. 
	We also spatially resolve the imprint of the entanglement on the one- and two-body density distributions showcasing that it accelerates the phase separation process or acts against spatial localization for repulsive and attractive intercomponent interactions respectively. 
	The many-body effects on the tunneling dynamics of the individual components, resulting from their counterflow, are also discussed. 
	The tunneling period of the impurity is very sensitive to the value of the impurity-medium coupling due to its effective dressing by the few-body medium. 
	Our work provides implications for engineering localized structures in correlated impurity settings using species selective optical potentials.
	
\end{abstract}

\maketitle

\section{Introduction}\label{Intro} 

Multicomponent quantum gases can be experimentally studied with a high degree of controllability in the ultracold regime~\cite{bloch2008many,bloch2017probing}. 
Specifically, two-component mixtures of bosons or fermions can be trapped in various species selective external geometries~\cite{fukuhara2009all,henderson2009experimental}. 
Few-body ensembles can be realized in particular in one-dimension (1D)~\cite{serwane2011deterministic,schmied2016bell} while the scattering lengths are tunable through Feshbach and confinement induced resonances~\cite{chin2010feshbach,kohler2006production}. 
In 1D bosonic mixtures the adjustability of the intercomponent interactions gives rise to intriguing phenomena such as phase-separation processes~\cite{mistakidis2018correlation,pyzh2020phase} in the repulsive regime, formation of bound states, e.g., droplet configurations~\cite{petrov2016ultradilute,parisi2020quantum} for attractive interactions as well as quasiparticle-like states in highly particle imbalanced systems~\cite{catani2012quantum,meinert2017bloch}. 

In this latter context, an impurity species is embedded in an environment 
of the majority species called the medium. 
The presence of a finite impurity-medium coupling leads to an effective picture where the impurity properties deviate from the bare particle case exhibiting, for instance, an effective mass~\cite{mistakidis2019effective,ardila2015impurity,grusdt2017bose,tajima2018many} and induced interactions~\cite{dehkharghani2018coalescence,mistakidis2020induced,takahashi2020extracting,brauneis2021impurities} mediated by the medium. 
The resultant states are often called polarons~\cite{massignan2014polarons,schmidt2018universal} and have been experimentally realized mainly in higher-dimensions~\cite{fukuhara2013u,yan2020bose,scazza2017repulsive,jorgensen2016observation,cetina2016ultrafast} and to a lesser extent in 1D~\cite{catani2012quantum,wenz2013few} using spectroscopic schemes. 
Since these settings consist of a few-body subsystem they naturally show enhanced correlation properties, especially in 1D, rendering their many-body treatment inevitable. 
In particular, the emergent impurity-medium entanglement can lead to spatial undulations of the medium. 
This mechanism is manifested, for instance, as sound-wave emission~\cite{mistakidis2020radiofrequency,grusdt2017bose} and collective excitations~\cite{mistakidis2020many,boudjemaa2020breathing} of the host or the formation of a bound state~\cite{ardila2020strong,camacho2018bipolarons,mukherjee2020induced} between the impurity and atoms of the medium for attractive interspecies interactions. 

Another relevant ingredient is the external trapping geometry that the two components experience. 
Indeed, for harmonically trapped and homogeneous systems remarkable dynamical features of impurity physics include the spontaneous generation of localized patterns~\cite{bougas2020pattern,grusdt2017bose,tajima2020collisional}, inelastic collisional aspects of driven impurities~\cite{mistakidis2019dissipative,mukherjee2020pulse,theel2020many} with the surrounding and their relaxation at long timescales~\cite{mistakidis2020pump,lausch2018prethermalization,palzer2009quantum}. 
On the other hand, when a lattice potential is introduced the situation becomes more complicated giving rise, among others, to doped insulator physics~\cite{keiler2020doping,bohrdt2020dynamical} and impurity transport~\cite{cai2010interaction,johnson2011impurity,theel2020entanglement}. 
Apparently, configuring one component by manipulating its external trap while leaving the other intact, e.g. by using a species selective external potential, it is possible to control the response of the unperturbed component via the impurity-medium interaction~\cite{keiler2018state,keiler2018correlation}. 
For instance, operating in the lowest-band approximation it has been demonstrated that a lattice trapped impurity interacting with a homogeneous host exhibits besides tunneling dynamics~\cite{keiler2019interaction} also self-trapping events~\cite{bruderer2008self,yin2015polaronic} and can even undergo Bloch-oscillations~\cite{grusdt2014bloch}. 
The opposite case, where the medium resides in the lattice, provides an experimental probe of the impurity-medium collision parameters~\cite{weber2010single} and interaction strength~\cite{will2011coherent}.  

In this work by considering an impurity in a box potential and a lattice trapped few-body medium we examine how the latter affects the impurity's spatial distribution by means of (de-)localization for different lattice depths and intercomponent interactions. 
Indeed, a lattice trapped medium can reside either in a superfluid or an insulating-like phase~\cite{keiler2020doping}, a fact that is expected to crucially impact the impurity's configuration and vice versa~\cite{tajima2021polaron}. 
To address the ground state properties and quantum quench dynamics of the above-discussed impurity setting we utilize the multi-layer multi-configuration time-dependent Hartree method for atomic mixtures (ML-MCTDHX)~\cite{cao2017unified,cao2013multi,kronke2013non}. 
This variational method enables us to account for the relevant correlations of the mixture and operate beyond the lowest-band approximation for the medium. 

Focusing on the ground state of the system and in order to testify its overall response for varying intercomponent interactions we determine the fidelity between the coupled and decoupled composite system both at the many-body and the single-particle level. 
Note that in impurity settings this observable is commonly termed residue~\cite{massignan2014polarons,schmidt2018universal} enabling us to identify e.g. the polaron formation, while the influence of the impurity-medium entanglement in this observable is still an open issue. 
It is demonstrated that despite the fact that the total entangled state may strongly deviate from its decoupled configuration, this effect is arguably less pronounced or even diminished at the single-particle level. 
Furthermore, we showcase that the build-up of impurity-medium entanglement is sensitive to the interplay between the intercomponent interactions and the lattice depth~\cite{keiler2020doping}. 
Interestingly, stronger interactions do not necessarily lead to a larger amount of entanglement, whereas the state of the majority species may undergo substantial structural changes, which remain invisible at the single-particle level. 
Moreover, we identify the imprint of the background on the impurities and vice versa by relying on one- and two-body density distributions evincing a rich spatial structure of the components with respect to the lattice depth as well as the inter- and intracomponent interactions. 
In particular, it is argued that for repulsive (attractive) interactions the impurity delocalizes (localizes) around the central lattice site. 
The delocalization of the impurity is accompanied by its phase-separation with the majority component~\cite{mistakidis2019quench}, where the impurity tends to the edges of the box for a superfluid background or exhibits a multi-hump structure for an insulating medium. 
We further analyze how much the intercomponent correlations are actually involved
in the structural changes observed in the spatial probability distributions.
To this end we compare density distributions of the numerically exact ground state 
to the corresponding ones of an approximate non-entangled ground state.
We identify that the entanglement-induced corrections 
accelerate phase-separation at repulsive couplings 
and generally slow down spatial localization at attractive interactions. 

Finally, we monitor the non-equilibrium dynamics of the mixture. 
We prepare the system in a phase-separated, i.e. disentangled configuration, and quench the intercomponent interactions to smaller values resulting in the counterflow of the components and thus triggering their tunneling dynamics and the consequent build-up of entanglement. 
The majority component plays the role of a material barrier for the impurity~\cite{theel2020entanglement,pflanzer2009material} which performs tunneling oscillations  
whose period depends strongly on the impurity-medium interaction. 
The many-body nature of the tunneling process of the components is testified by invoking the individual natural orbitals constituting the time-evolved many-body state. 

Our presentation is structured as follows. 
In \cref{sec:setup} we introduce the impurity setting 
and in \cref{sec:methodology} we discuss our 
many-body treatment to tackle its ground state and dynamics. 
The ground state properties of the delocalized impurity 
and the lattice trapped medium are addressed in \cref{sec:ground_state}. 
We analyze the fidelity between perturbed and unperturbed (reduced) density operators, quantify the degree of entanglement and visualize its impact on single- and two-body density distributions of each species for different intra- and intercomponent interactions and lattice depths. 
The non-equilibrium dynamics of the mixture following a quench of the impurity-medium coupling to smaller values is discussed in \cref{sec:dyn}. 
We provide a summary of our results and elaborate on future perspectives in \cref{sec:summary}.

\section{Setup and Hamiltonian}
\label{sec:setup}

We consider a single impurity particle 
immersed in a few-body system of ultracold bosons.
Both components reside in a quasi-1D geometry ensured by a strong transversal confinement~\cite{catani2012quantum}.
Along the longitudinal direction the $N_A$ majority species atoms of mass $m_A$ are trapped 
inside a lattice of depth $V$ with $l$ sites and length $L$ with hard-wall boundary conditions.
The impurity atom of mass $m_B$ is subject to a box potential of the same length.
The species-dependent trapping has been successfully demonstrated experimentally ~\cite{fukuhara2009all,henderson2009experimental}.
The inter-particle interactions are of s-wave contact type with
$g_{AA}$ denoting the majority-majority interaction strength 
and $g_{AB}$ the majority-impurity coupling.
Both may be tuned independently by a combination of Feshbach 
and confinement induced resonances~\cite{chin2010feshbach,kohler2006production}.
Furthermore, we assume equal masses $m_A=m_B$, which corresponds to a mixture
of the same isotope with the particles being distinguishable 
due to two different hyperfine states~\cite{Myatt1997,Hall1998,Ketterle1999,Inguscio2000,Hall2007,Becker2008}.
By introducing $R^*=L$ and $E^*=\hbar^2/(m L^2)$ as length and energy scales 
we arrive at the following rescaled many-body Hamiltonian:
\begin{equation}
\begin{aligned}
\label{eq:hamilttt}
	H =& - \frac{1}{2} \frac{\partial^2}{\partial y^2}
	- \sum_{i}^{N_{A}} \left( \frac{1}{2} \frac{\partial^2}{\partial x_i^2} + V \sin^{2}(\pi l x_i) \right)\\ 
	&+ g_{AA} \sum_{i<j}^{N_A} \delta(x_i-x_j) + g_{AB} \sum_{i}^{N_A} \delta(x_i-y),
\end{aligned}
\end{equation}
where $y$ and $x_i$ denote the spatial coordinates of the impurity 
and {\it i}th majority atom respectively.  

In this work we primarily focus on the ground state properties 
of the above many-body Hamiltonian \cref{eq:hamilttt} with $l=5$ lattice sites 
and $N_A=5$ majority particles.
In particular, we are interested in the susceptibility of the composite system
to structural changes and the amount of inter-particle correlations it may hold.
We cover a parameter space from moderately attractive to repulsive interaction strengths, i.e.,
$g_{AA} \in \left[- 3.0, 3.0 \right] E^* R^*$ 
and $g_{AB} \in \left[- 5.0, 5.0 \right] E^* R^*$,
for a  range of lattice depths from shallow to deep, namely 
$V \in \left[100, 1000 \right] E^*$. 
In the following, we will refer to a lattice as being shallow ($V<200$), 
moderately deep ($V\approx500$) and very deep ($V>800$).
Additionally, we demonstrate how an initially disentangled state 
prepared in the immiscible regime
acquires dynamically a finite amount of entanglement
after quenching the intercomponent coupling $g_{AB}$, thus
triggering a counter-flow tunneling process of the two components.

% % % % % % % % % % % % % % % % % % % % % % % % % % % % % % % % %
% % % % % % % % % % % % % % % % % % % % % % % % % % % % % % % % %
\section{Variational Approach}
\label{sec:methodology}

In order to account for effects stemming 
from inter-particle correlations
we rely on the Multi-Layer Multi-Configurational 
Time-Dependent Hartree Method for atomic mixtures (ML-MCTDHX), for short ML-X~\cite{cao2017unified,cao2013multi,kronke2013non}.
This ab-initio method has been successfully applied 
to solve the time-dependent Schrödinger equation
of various experimentally accessible and extensively studied systems.
The core idea of this method lies
in expanding the many-body wave-function 
in terms of product states of {\it time-dependent} single-particle functions~\cite{lode2020colloquium,alon2021many}.
This becomes beneficial, when the number of basis configurations
with considerable contribution to the state fluctuates weakly
during the time propagation, whereas the configurations themselves do change. 
Taking a variationally optimal basis at each time-step 
allows us to cover the high-dimensional Hilbert space
at a lower computational cost compared to a time-independent basis.

The wave function ansatz for a given system 
is decomposed in multiple layers. 
On the first layer, called top layer, 
we separate the degrees of freedom of the binary mixture 
into product states of majority
and impurity species functions $\ket{\Psi_i^{\sigma}(t)}$ 
with $\sigma \in \{A,B\}$ and $i \in \{1, \ldots, S\}$:
\begin{equation}
\ket{\Psi(t)} =\sum_{i=1}^{S}\sqrt{\lambda_i(t)} 
\ket{\Psi_i^A(t)} \otimes \ket{\Psi_i^B(t)}.
\label{eq:wfn_ansatz_species_layer}
\end{equation}
Here, the time-dependent coefficients $\lambda_i(t)$, 
normalized as $\sum_{i=1}^{S} \lambda_i(t) = 1$, 
determine the degree of entanglement between the components~\cite{horodecki2009quantum}.
The choice of $S=1$
results in the so-called species mean-field (SMF) approximation, 
meaning that no entanglement is assumed between the components~\cite{mistakidis2019effective}.
In that case the intercomponent correlations, if present, are neglected
and every component is effectively subject to an additional one-body potential
induced by the fellow species~\cite{mistakidis2019quench,theel2020entanglement}. 
In this work, we put a special emphasis on the impact of the entanglement
on several one- and two-body quantities by comparing 
the numerically exact ground state to the corresponding SMF approximation.

On the second layer, called species layer, each species function $\ket{\Psi_i^{\sigma}(t)}$
is expanded in terms of species-dependent symmetrized product states
of single-particle functions (SPFs) $\ket{\varphi_{j}^{\sigma}(t)}$ 
with $j \in \{1, \ldots, s_{\sigma}\}$,
accounting for the bosonic nature of our particles
and abbreviated as $\ket{\vec{n}^{\sigma}}=\ket{n_1^{\sigma},\ldots, n_{s_{\sigma}}^{\sigma}}$:
\begin{equation}
\ket{\Psi_i^{\sigma}(t)} = \sum_{\vec{n}^{\sigma}|N_\sigma}
C_{i, \vec{n}^{\sigma}}(t)
\ket{\vec{n}^{\sigma}(t)}.
\label{eq:wfn_ansatz_particle_layer}
\end{equation}
In this expression the sum is performed over all configurations $\vec{n}^{\sigma}|N_\sigma$
obeying the particle-number constraint 
$\sum_{i=1}^{s_{\sigma}} n_i^{\sigma} = N_{\sigma}$.
On the third and final layer, called primitive layer, each SPF is represented 
on a one-dimensional time-independent grid~\cite{DVR1985}.

The Dirac-Frenkel variational principle~\cite{DiracFrenkel2000} is subsequently applied 
to the above ansatz in order to derive the
coupled equations of motion for the expansion coefficients 
$\lambda_i(t)$, $C_{i, \vec{n}^{\sigma}}(t)$ 
and the SPFs $\ket{\varphi_{j}^{\sigma}(t)}$.
Finally, performing imaginary time-evolution one arrives 
at the ground state wave-function (\cref{sec:ground_state}),
whereas the real time-propagation
allows to study the non-equilibrium dynamics
of an arbitrary initial state (\cref{sec:dyn}).
The results to be presented below have been obtained by using
$(S,s_A,s_B)=(4,5,4)$ functions/SPFs on the
top/species layers as well as $225$ grid points on the primitive layer.
We have carefully checked the convergence behavior of our results 
by comparing to simulations
with a larger number of orbitals $(S,s_A,s_B)=(6,8,6)$ 
and found no significant changes for the quantities of interest.

In the following we will often refer to the 
reduced $j$-body density operators $\hat{\rho}_j^{\sigma}$ 
of species $\sigma$ and the
intercomponent reduced $(j+k)$-body density operator $\hat{\rho}_{j+k}^{\sigma \bar{\sigma}}$ 
obtained from the many-body density operator $\hat{\rho} = \ket{\Psi}\bra{\Psi}$:
\begin{flalign}
\hat{\rho}_j^{\sigma} = \tr_{_{N_{\sigma} \setminus j}}
\{\tr_{_{N_{\bar{\sigma}}}} \{{\hat{\rho}}\}\} \label{eq:rho1},\\
\hat{\rho}_{j+k}^{\sigma \bar{\sigma}} = 
\tr_{_{N_{\sigma} \setminus j}}\{\tr_{_{N_{\bar{\sigma}} \setminus k}} 
\{\hat{\rho}\}\} ,\label{eq:rho2inter}
\end{flalign}
where $N_{\sigma} \setminus j$ stands for integrating out 
$N_{\sigma}-j$ coordinates of component $\sigma$ and $\bar{\sigma} \neq \sigma$.
Of particular interest are the reduced one-body density operators 
$\hat{\rho}_1^{A}$ and $\hat{\rho}_1^{B}$
as well as the reduced two-body intra- and intercomponent density operators 
$\hat{\rho}_2^{A}$ and $\hat{\rho}_2^{AB}$ respectively, 
since they determine the expectation values of various experimentally accessible 
local one- and two-body observables, 
such as the average particle position, the inter-atomic distance or the wave-packet width.

% % % % % % % % % % % % % % % % % % % % % % % % % % % % % % % % %
% % % % % % % % % % % % % % % % % % % % % % % % % % % % % % % % %
\section{Impact of intercomponent coupling on Ground state properties}
\label{sec:ground_state}

In \cref{subsec:fidelity} we analyze to which extent
the many-body wave-function as well as the reduced
one-body density operators 
are modified by the intercomponent interaction.
To this end we analyze the fidelity 
between the interacting and non-interacting (reduced) density operators, 
which is a measure of their closeness.
We find that with increasing absolute value of the interaction strength 
the system is more robust w.r.t.\ changes
on the one-body as compared to the many-body level. 
Moreover, each component is affected differently 
depending on the lattice depth and majority interaction strength.

Subsequently, in \cref{subsec:entropy} 
we quantify the degree of entanglement by means of the von-Neumann entropy 
and identify parameter regions with substantial inter-particle correlations.
Interestingly, increasing the absolute value of the intercomponent coupling does not always
result in stronger entanglement. 
In fact, there are parameter regions where
a strongly interacting ground state becomes almost orthogonal
to the non-interacting one and the components remain to a good approximation disentangled.

Finally, we combine insights from \cref{subsec:fidelity,subsec:entropy} to identify
interesting parameter regimes 
and perform an in-depth analysis of the underlying physical phenomena in \cref{subsec:dmat}.
In particular, we inspect how the spatial representation of density operators is altered
and compare those to the corresponding SMF results.
The latter allows us to spatially resolve the corrections to the SMF densities 
induced by the entanglement
and interpret its impact as acceleration or deceleration of the undergoing processes,
e.g., the phase separation or localization.

% % % % % % % % % % % % % % % % % % % % % % % % % % % % % % % % %
\subsection{Fidelity for quantifying the impact of the intercomponent interaction}
\label{subsec:fidelity}

First, we aim to analyze how the intercomponent coupling $g_{AB}$
impacts the ground state of non-interacting species (NIS) at $g_{AB}=0$.
For this purpose, we evaluate the fidelity \cite{fidelity1994}
of two density operators $\hat{\rho}$ and $\hat{\sigma}$ defined as:

\begin{equation}
\label{eq:fidelity}
	F(\hat{\rho}, \hat{\sigma}) 
	= \left(
	\tr \sqrt{\sqrt{\hat{\rho}}\hat{\sigma}\sqrt{\hat{\rho}}}
	\right)^2 
	= F(\hat{\sigma}, \hat{\rho}).
\end{equation}

\begin{figure*}[t]
	{\includegraphics[width=0.31\textwidth]{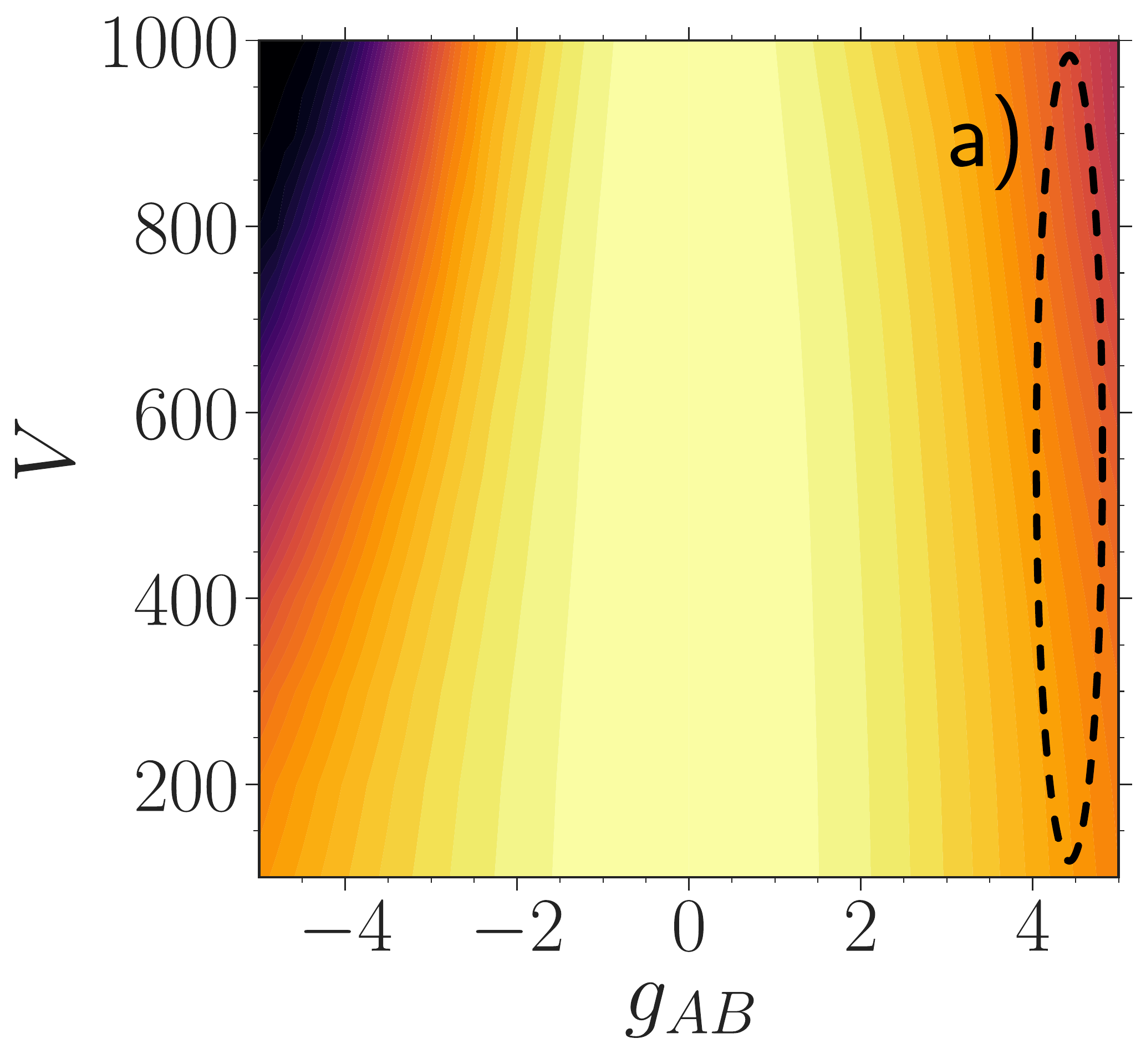}}
	{\includegraphics[width=0.30\linewidth]{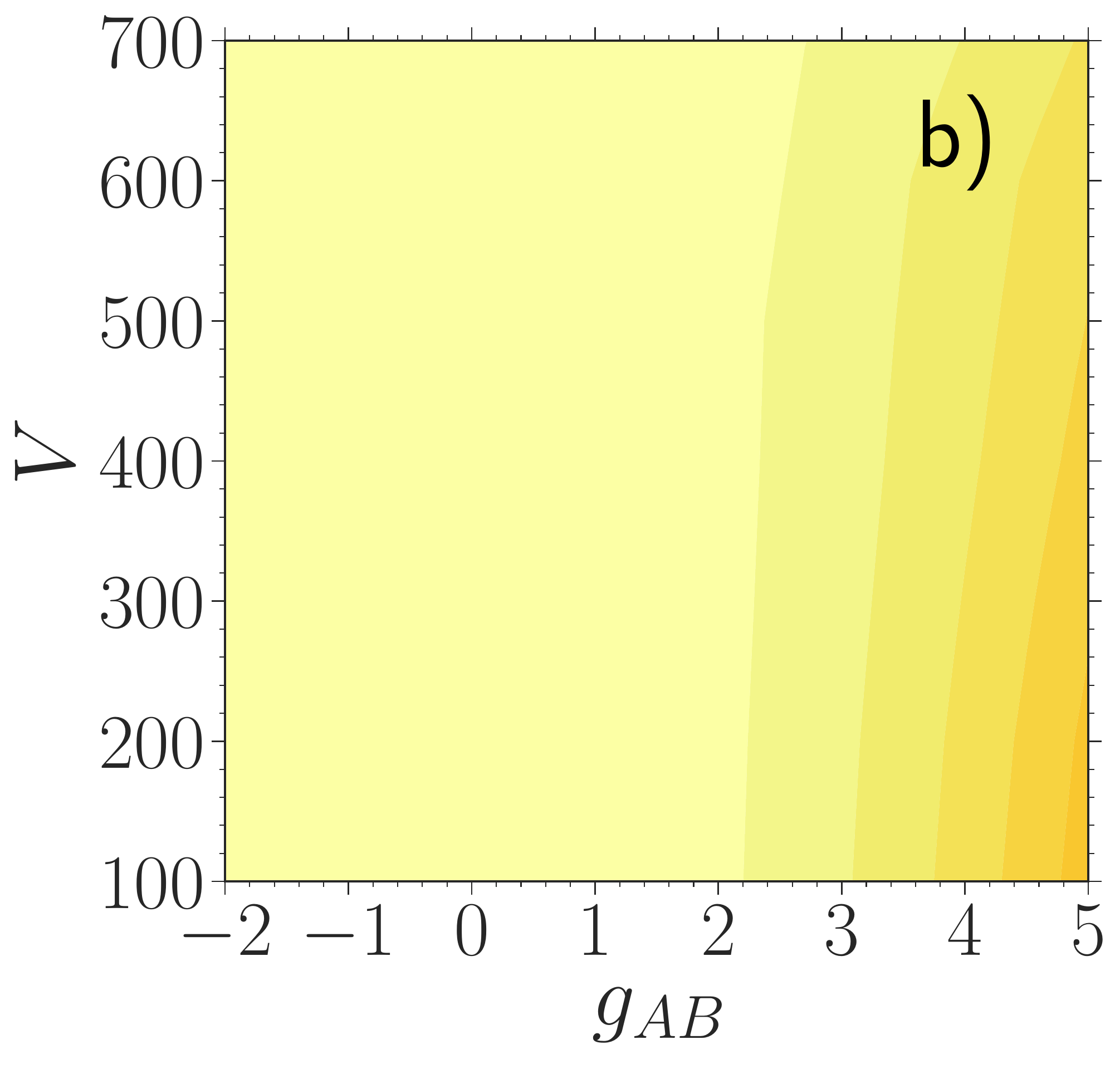}}
	{\includegraphics[width=0.37\linewidth]{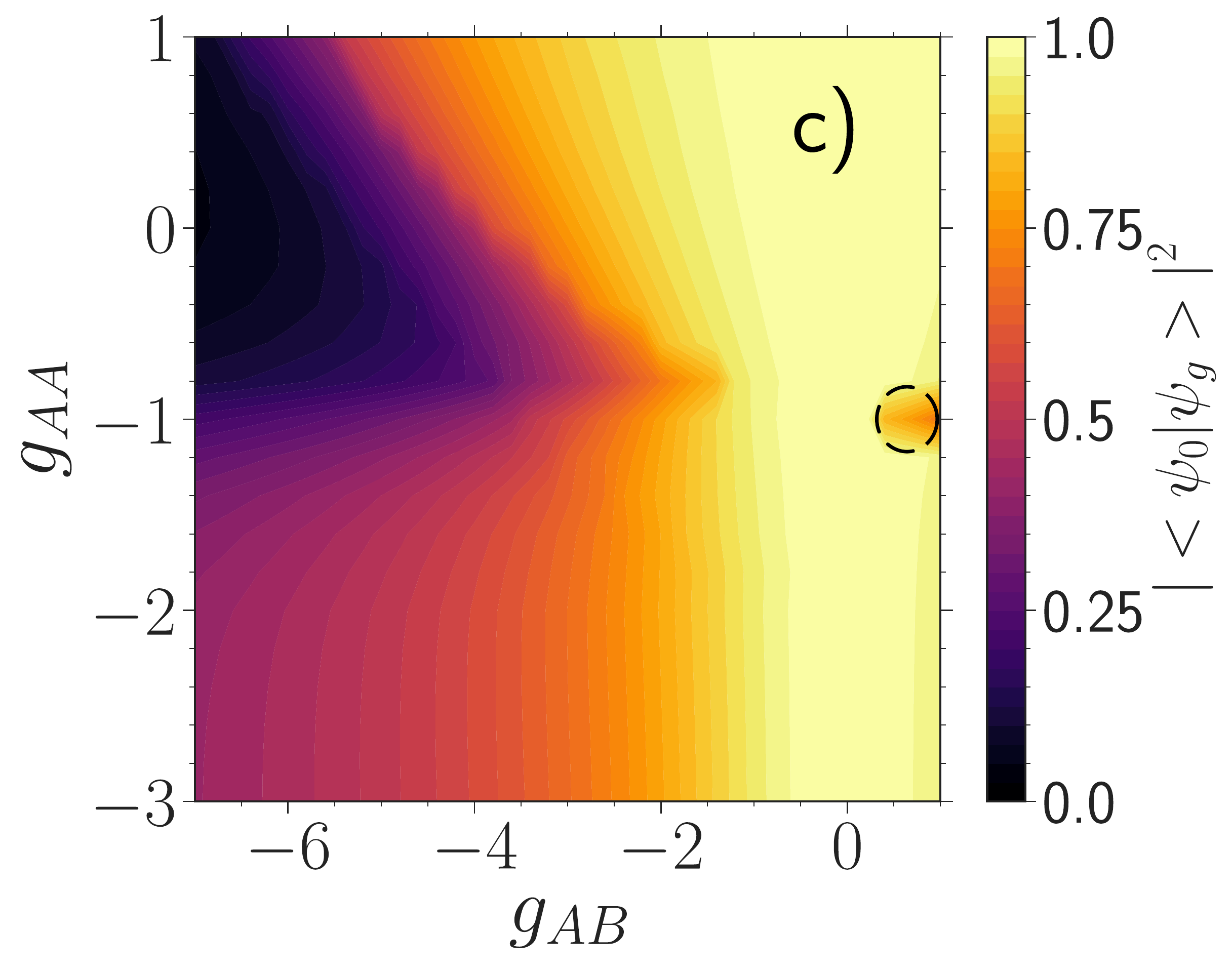}}
%	\begin{subfigure}{.31\textwidth}
%		\includegraphics[width=1\linewidth]{res_gAA05.pdf}
%	\end{subfigure}%
%	\begin{subfigure}{.30\textwidth}
%		\includegraphics[width=1\linewidth]{res_gAA30.pdf}
%	\end{subfigure}
%	\begin{subfigure}{.37\textwidth}
%		\includegraphics[width=1\linewidth]{res_V0500.pdf}
%	\end{subfigure}
%	\begin{subfigure}{.066\textwidth}
%		\centering
%		\includegraphics[width=\linewidth]{res_cbar.pdf}
%	\end{subfigure}
	\caption{Fidelity 
		$|\braket{\Psi_0|\Psi_g}|^2$ 
		between a many-body state  
		$\ket{\Psi_0}$ at $g_{AB}=0$ and a many-body one $\ket{\Psi_g}$ at finite $g_{AB}$,	
		for a) $g_{AA}=0.5$, b) $g_{AA}=3.0$ and c) $V=500$
		as a function of the majority-impurity coupling $g_{AB}$ 
		and the lattice depth $V$ (a,b) 
		or the interaction strength of the majority atoms $g_{AA}$ (c).
		All quantities are given in box units with characteristic length 
		$R^*=L$ and energy $E^*=\hbar^2/(m L^2)$ 
		with $L$ denoting the extension of the box trap.
		Regions encircled by black dashed lines indicate parameter regions with
		substantial qualitative differences to the SMF ansatz.
		}
	\label{fig:fidelity_full}
\end{figure*}

We start with the fidelity between 
a NIS many-body density 
$\hat{\rho}_0 = \ket{\Psi_0}\bra{\Psi_0}$
and a many-body density $\hat{\rho}_g=\ket{\Psi_g}\bra{\Psi_g}$ 
for some finite coupling $g_{AB}$ (\cref{fig:fidelity_full}).
Since both density operators describe pure states, 
\cref{eq:fidelity} reduces to 
$F_{mb} = |\braket{ \Psi_0 | \Psi_g}|^2$.
This measure, $F_{mb}$, is also known as the polaron residue
studied in the context of phonon dressing of an impurity particle
immersed in a bath of majority atoms~\cite{massignan2014polarons,schmidt2018universal}.

For a weakly interacting ($g_{AA}=0.5$) majority component
[\cref{fig:fidelity_full}a] we observe that the many-body fidelity
at a fixed lattice depth
decreases monotonously with the modulus of the coupling strength $g_{AB}$.
At deep lattices the rate of its reduction is larger, 
a behavior which is even more pronounced at
strong negative $g_{AB}$, where the interacting state 
becomes almost orthogonal to the non-interacting one ($g_{AB}=-5$ and $V=1000$).
The black dashed line encircles a parameter region of instability 
where the SMF ansatz 
collapses to a configuration with broken parity symmetry.
For a moderately interacting ($g_{AA}=3.0$) majority component 
[\cref{fig:fidelity_full}b] the many-body fidelity becomes much more stable.
Contrarily to \cref{fig:fidelity_full}a
the rate of reduction with $g_{AB}$ is larger
at shallow lattices instead.
Finally, for a moderately deep ($V=500$) lattice [\cref{fig:fidelity_full}c]
we observe a peculiar fast decay 
around $g_{AA} \approx -1$ starting at $g_{AB}<-2$.
Additionally, at $g_{AA}\approx-1$ and positive $g_{AB}$ 
there is a small pronounced decay region (black dashed circle),
which is absent in the SMF approximation.

\begin{figure*}[t]
	{\includegraphics[width=0.31\textwidth]{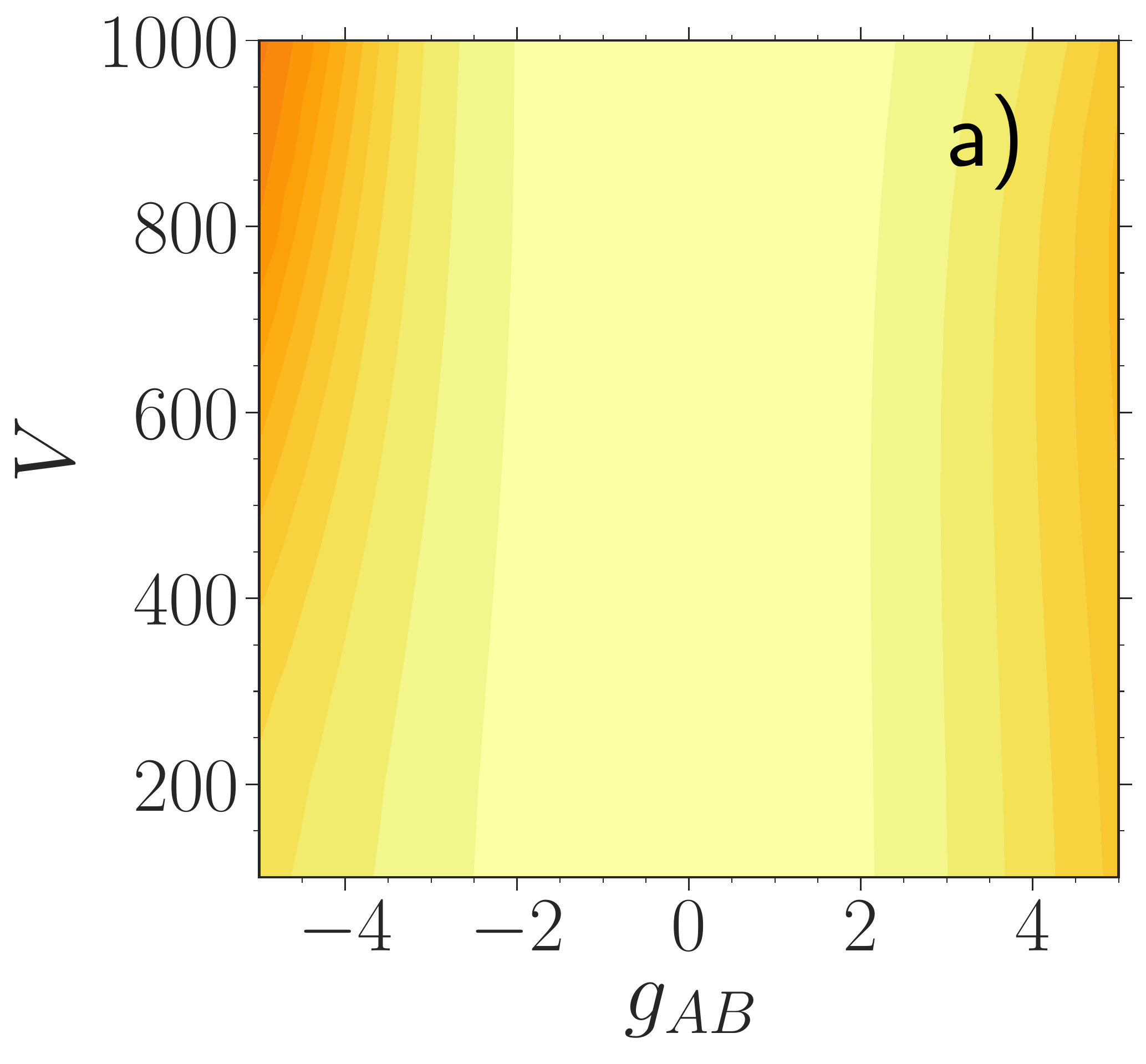}}
	{\includegraphics[width=0.30\linewidth]{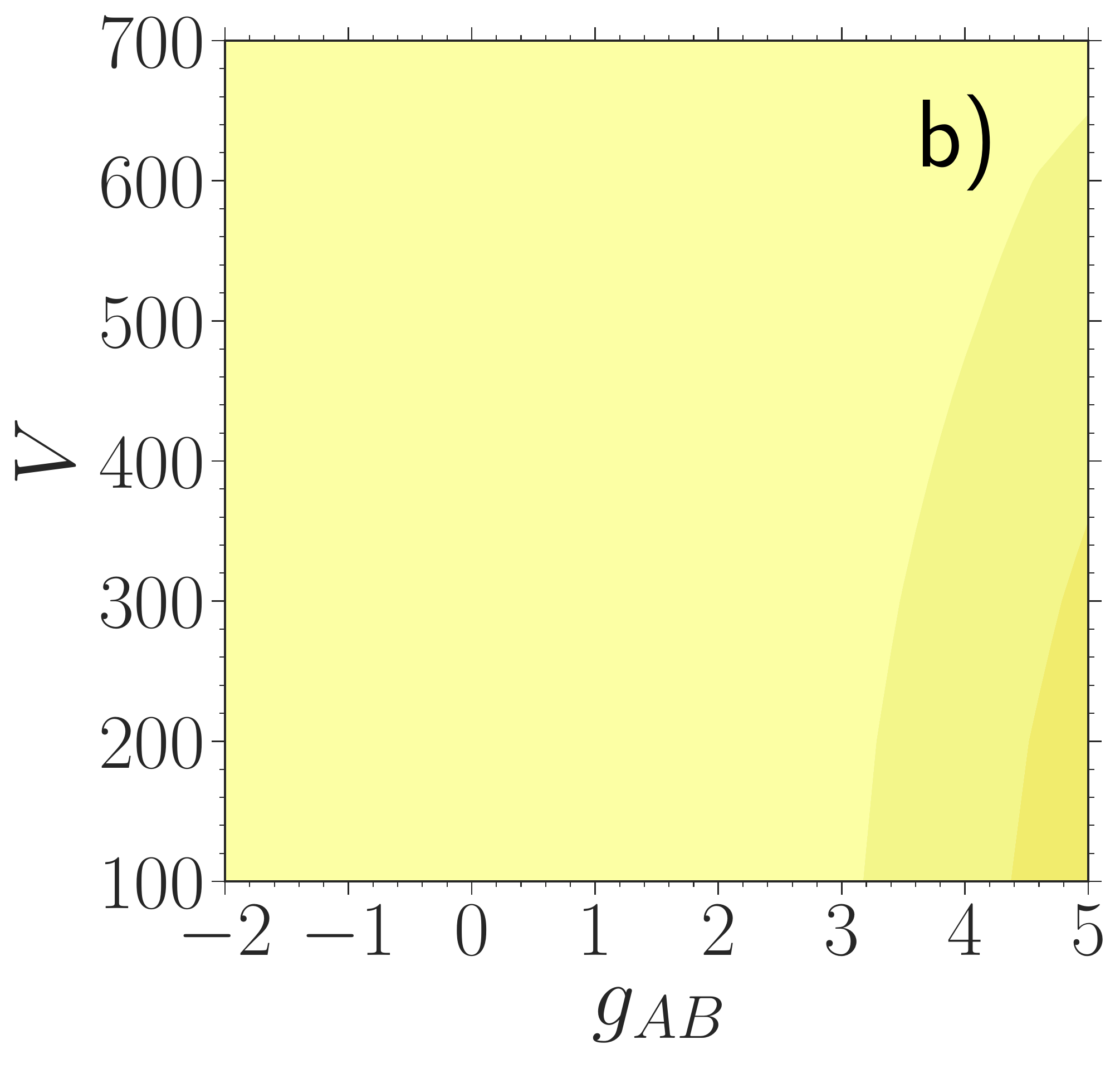}}
	{\includegraphics[width=0.37\linewidth]{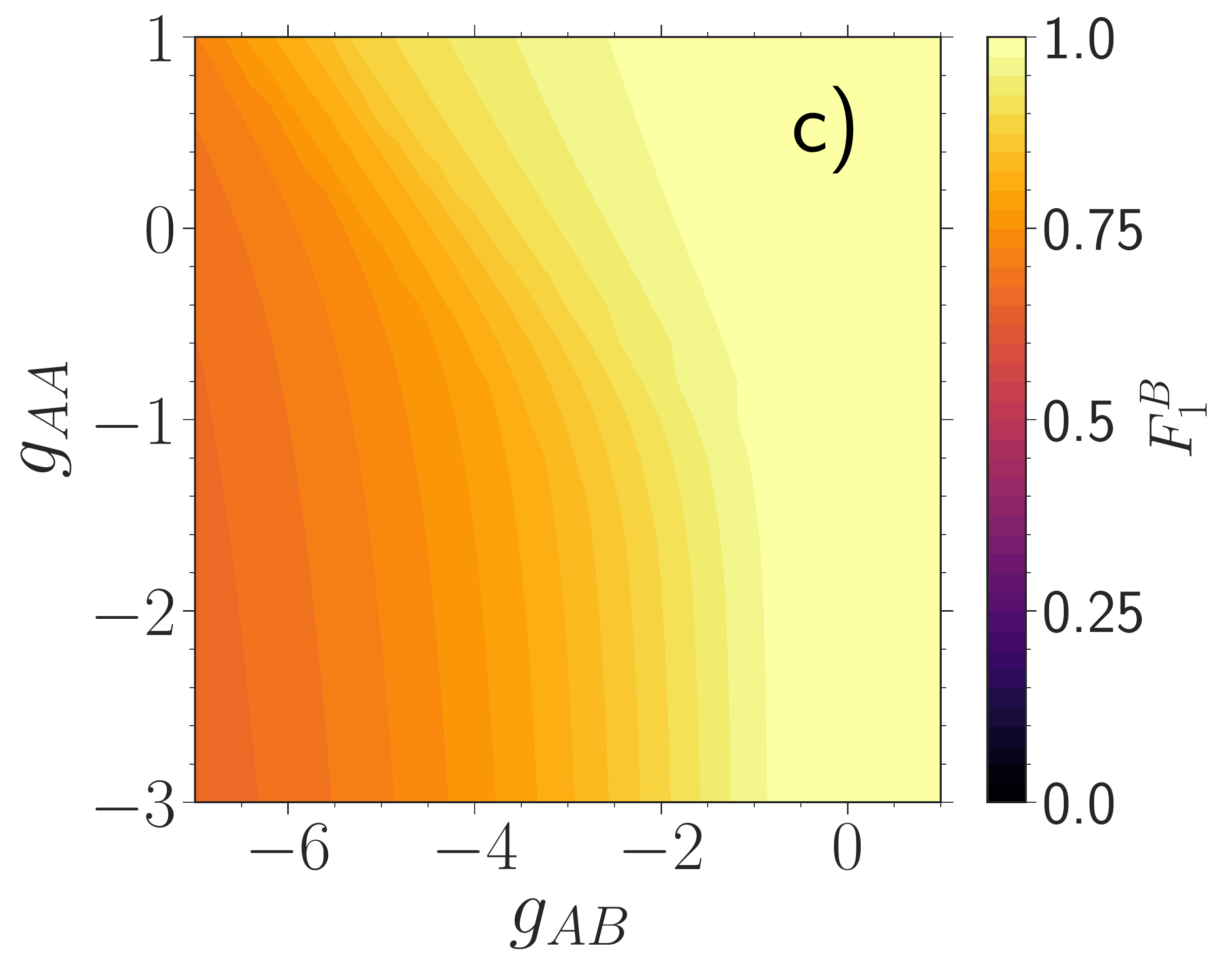}}
%	\begin{subfigure}{.31\textwidth}
%		\centering
%		\includegraphics[width=1\linewidth]{Uhlman_B_gAA05.pdf}
%	\end{subfigure}%
%	\begin{subfigure}{.30\textwidth}
%		\centering
%		\includegraphics[width=1\linewidth]{Uhlman_B_gAA30.pdf}
%	\end{subfigure}
%	\begin{subfigure}{.37\textwidth}
%		\centering
%		\includegraphics[width=1\linewidth]{Uhlman_B_V0500.pdf}
%	\end{subfigure}
	%	\begin{subfigure}{.066\textwidth}
	%		\centering
	%		\includegraphics[width=\linewidth]{Uhlman_B_cbar.pdf}
	%	\end{subfigure}
	\caption{Fidelity 
		$F_1^B = |\bra{ \Phi_0} \hat{\rho}_1^B \ket{\Phi_0}|^2$
		between a free impurity particle
		$\ket{\Phi_0}$ at $g_{AB}=0$ 
		and an entangled one $\hat{\rho}_1^B$ at finite $g_{AB}$,
		for a) $g_{AA}=0.5$, b) $g_{AA}=3.0$ and c) $V=500$
		and varying majority-impurity coupling $g_{AB}$ 
		and the lattice depth $V$ or the interaction strength of the majority atoms $g_{AA}$.
		All quantities are expressed in box units with characteristic length 
		$R^*=L$ and energy $E^*=\hbar^2/(m L^2)$ while 
		$L$ is the extension of the box trap.}
	\label{fig:fidelity_B}
\end{figure*}

Next, we analyze the fidelity between 
a free impurity described by a pure state $\ket{\Phi_0}\bra{\Phi_0}$ 
and an entangled one $\hat{\rho}_1^B$, 
in general being a mixed state [\cref{fig:fidelity_B}].
\cref{eq:fidelity} then simplifies to
$F_1^B = |\bra{ \Phi_0} \hat{\rho}_1^B \ket{\Phi_0}|^2$.
This measure allows to judge 
to which extent the impurity atom is still a "free" particle of mass $m_B$.
We emphasize that it should not be confused with
a polaron quasi-particle having a renormalized effective mass. 
We observe that $F_1^B$ follows overall 
a similar pattern as the many-body fidelity $F_{mb}$, 
but with a significantly slower decay rate. 
Though there are some strong qualitative differences, 
see in particular \cref{fig:fidelity_B}c.
Namely, the abrupt decay of $F_{mb}$ around $g_{AA} \approx -1$
at negative $g_{AB}$ [\cref{fig:fidelity_full}c] is absent in $F_1^B$ 
along with the small decay region at positive $g_{AB}$  (black dashed circle). 
From this we anticipate that the majority component is responsible 
for these features in $F_{mb}$.

\begin{figure*}[t]
	{\includegraphics[width=0.31\textwidth]{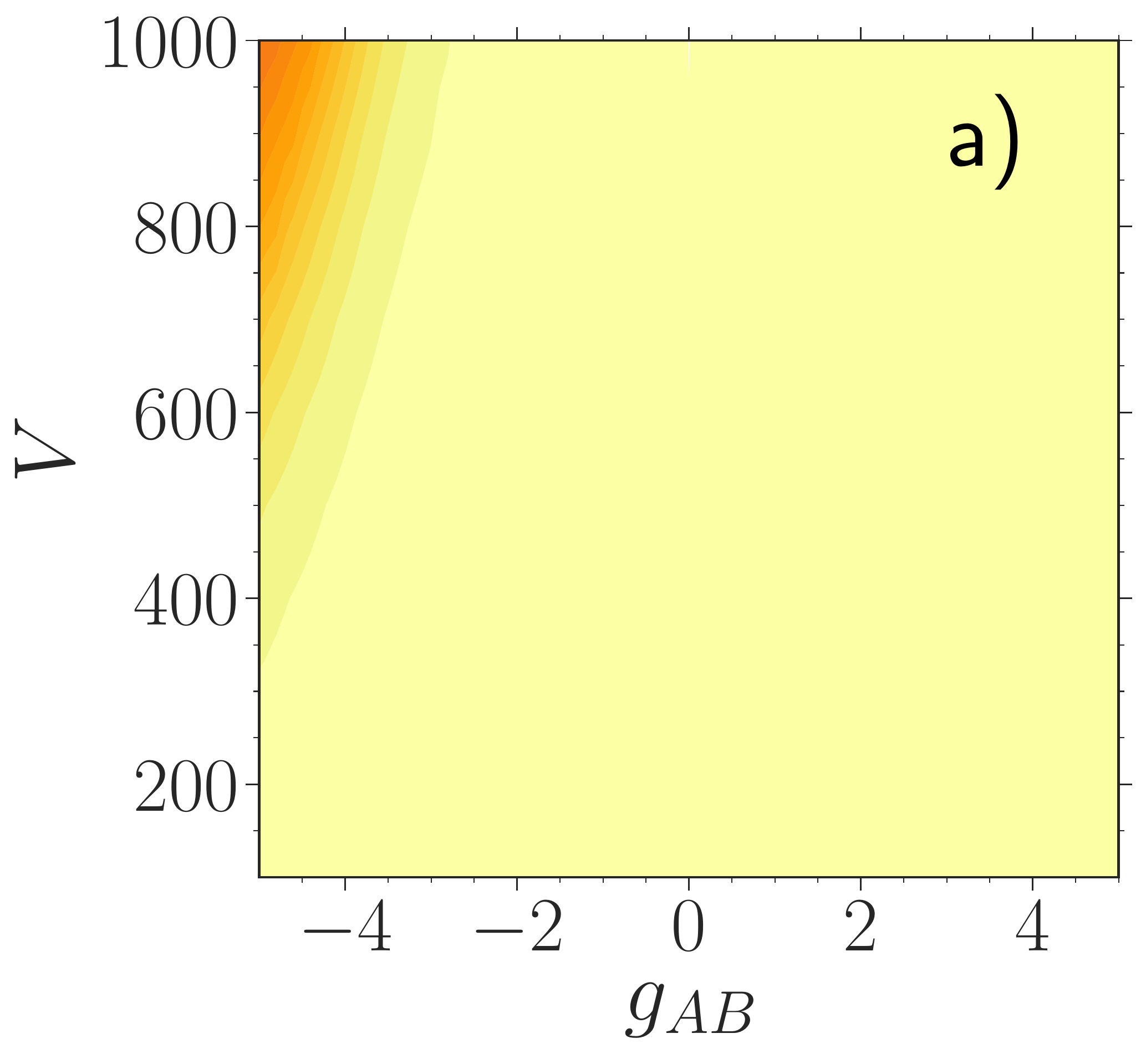}}
	{\includegraphics[width=0.30\linewidth]{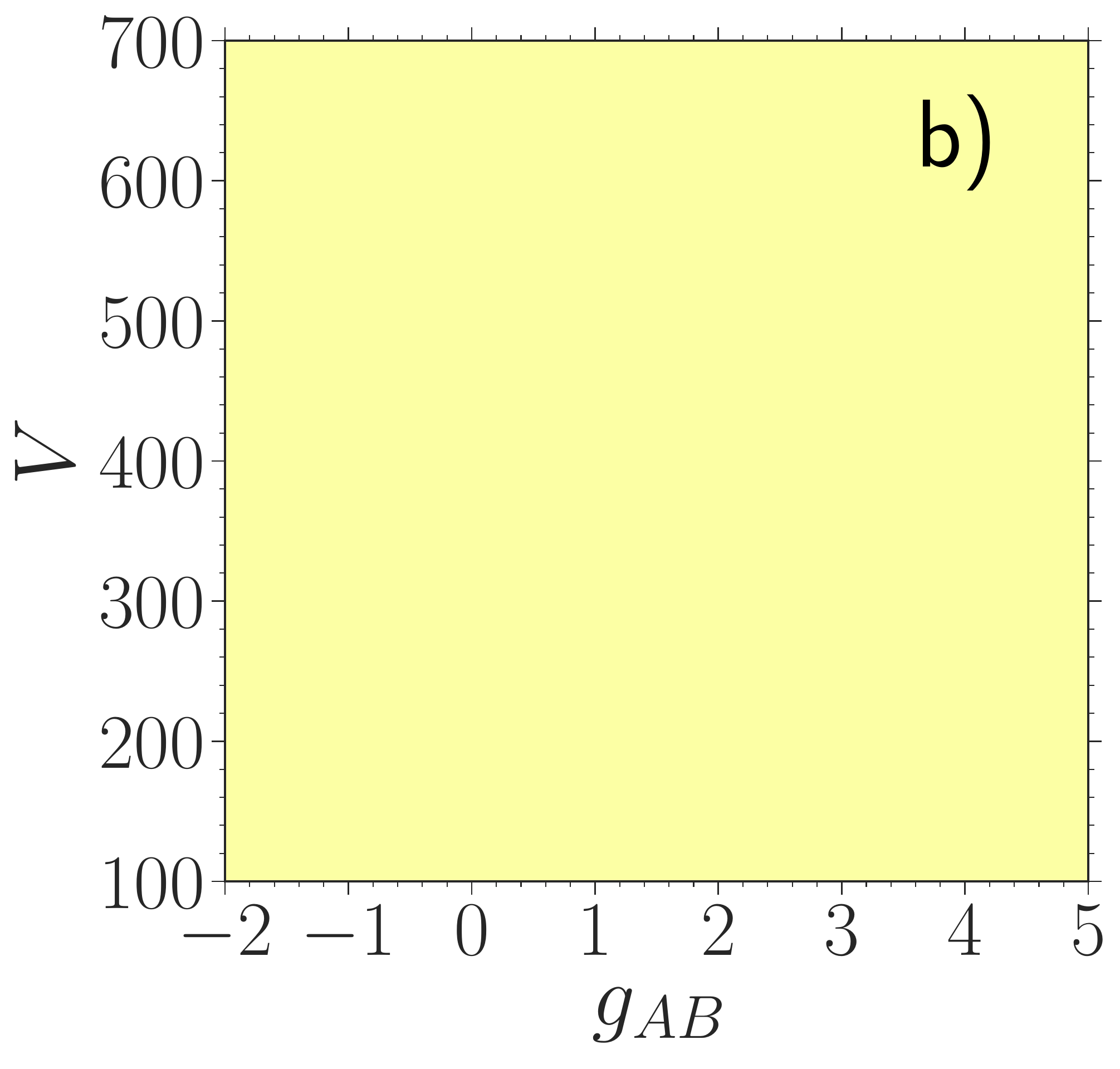}}
	{\includegraphics[width=0.37\linewidth]{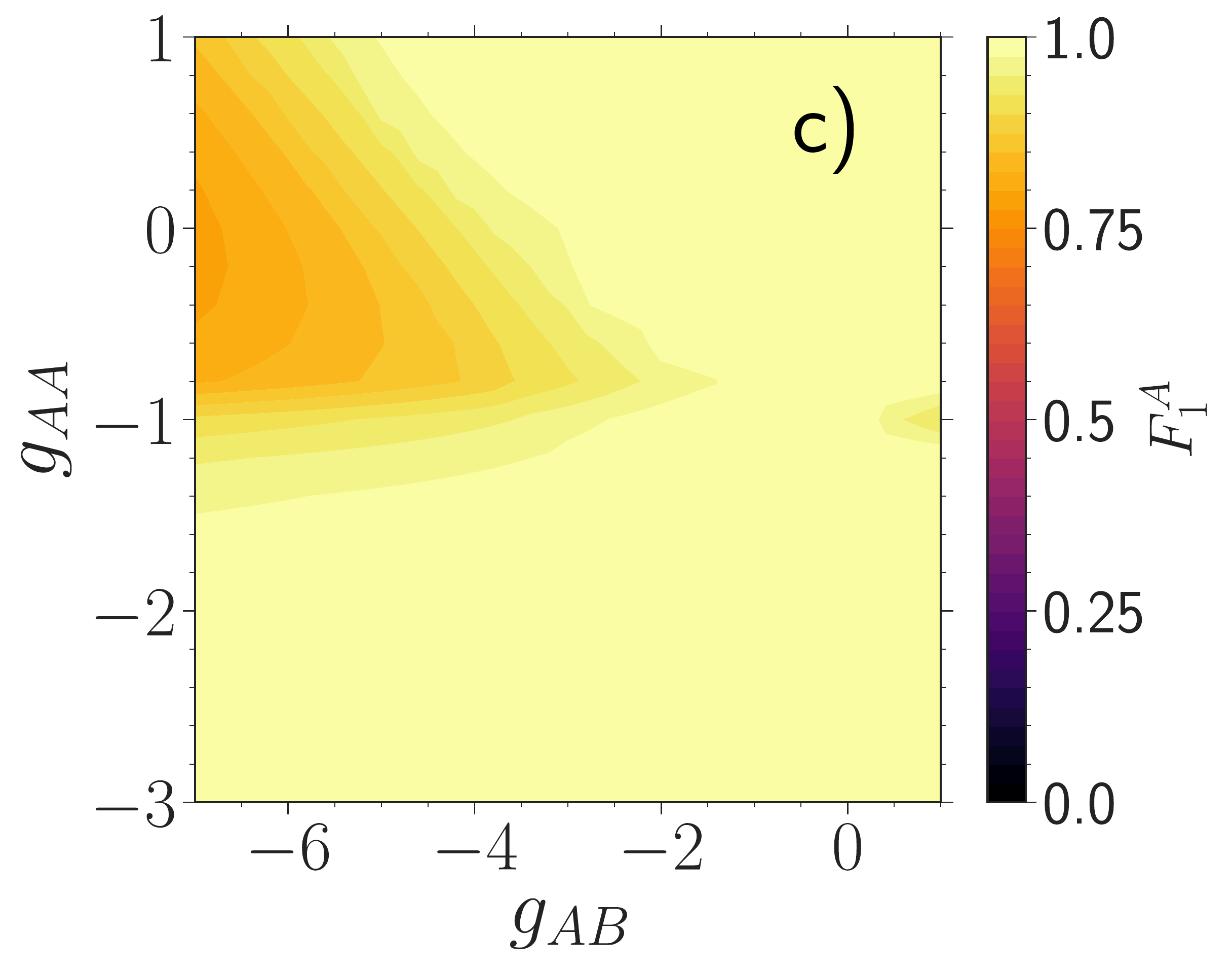}}	
%	\begin{subfigure}{.31\textwidth}
%		\centering
%		\includegraphics[width=1\linewidth]{Uhlman_A_gAA05.pdf}
%	\end{subfigure}%
%	\begin{subfigure}{.30\textwidth}
%		\centering
%		\includegraphics[width=1\linewidth]{Uhlman_A_gAA30.pdf}
%	\end{subfigure}
%	\begin{subfigure}{.37\textwidth}
%		\centering
%		\includegraphics[width=1\linewidth]{Uhlman_A_V0500.pdf}
%	\end{subfigure}
	%	\begin{subfigure}{.066\textwidth}
	%		\centering
	%		\includegraphics[width=\linewidth]{Uhlman_A_cbar.pdf}
	%	\end{subfigure}
	\caption{Fidelity 
		$F^A_1 = F(\hat{\rho}^A_1, \hat{\rho}^A_1(g_{AB}=0))$
		between mixed states
		characterizing a majority particle 
		when the medium is disentangled  $\hat{\rho}^A_1(g_{AB}=0)$ 
		and entangled $\hat{\rho}^A_1$ with the impurity atom,		
		for a) $g_{AA}=0.5$, b) $g_{AA}=3.0$ and c) $V=500$
		as a function of the majority-impurity coupling $g_{AB}$ 
		and the lattice depth $V$ (a,b) 
		or the interaction strength of the majority atoms $g_{AA}$ (c).
		All quantities are provided in box units of characteristic length 
		$R^*=L$ and energy $E^*=\hbar^2/(m L^2)$ 
		with $L$ being the extension of the box trap.}
	\label{fig:fidelity_A}
\end{figure*}

For the above reason, we now investigate the complementary fidelity 
$F_1^A = F(\hat{\rho}^A_1(g_{AB}=0),\hat{\rho}^A_1)$, i.e., 
between mixed states
characterizing a majority particle in the NIS state $\hat{\rho}^A_1(g_{AB}=0)$ 
and in the interacting state $\hat{\rho}^A_1$
[\cref{fig:fidelity_A}].
This quantity captures to which extent
a majority particle is still in a mixed state induced solely by
the intraspecies interaction strength $g_{AA}$.
In case of a weak $g_{AA}$ [\cref{fig:fidelity_A}a]
$F_1^A$ is notably affected only at deep lattices $V>600$
and strong negative coupling $g_{AB}<-4$.
For large $g_{AA}$ [\cref{fig:fidelity_A}b]
we observe that the intercomponent correlations 
are not strong enough to overcome the intraspecies ones,
thus barely affecting the mixedness of the NIS majority state, 
since $F_1^A \approx 1$ in the whole range $-2<g_{AB}<5$ and $100<V<700$.
In \cref{fig:fidelity_A}c we find evidence that
the majority component is indeed responsible for the particular decay patterns
observed in the many-body fidelity $F_{mb}$, which were absent in $F_1^B$. 
Overall, the majority component demonstrates a higher level of robustness at the single-particle level as compared to the impurity.

\newpage

% % % % % % % % % % % % % % % % % % % % % % % % % % % % % % % % %
\subsection{Entropy measures for quantifying the degree of correlations}
\label{subsec:entropy}

As we have seen in the previous section, 
an initially disentangled composite system may be drastically influenced 
by the intercomponent coupling. 
However, it is far from obvious to which extent the correlations 
are actually involved when the ground state undergoes structural changes~\cite{wang2016inter}.
For instance, 
a strongly interacting ground state may in fact just represent a different disentangled state
or a state seemingly unaffected by the coupling may feature substantial correlations 
which guarantee its robustness.
To investigate these intriguing possibilities
we perform a further classification based on the degree of inter-particle correlations.

To quantify the degree of correlations in our impurity system
we use the von-Neumann entropy of the reduced density operators~\cite{entropy2017}. 
Here, we distinguish between the {\it entanglement} entropy $S_{vN}$
of the reduced density operator $\hat{\rho}^{\sigma}$ of species $\sigma$~\cite{mistakidis2018correlation,keiler2020doping} and 
the {\it fragmentation} entropy $S_{vN}^{\sigma}$
of the reduced one-body density operator $\hat{\rho}_1^{\sigma}$ of species $\sigma$~\cite{roy2018phases,bera2020relaxation,keiler2018correlation}.
The former, $\hat{\rho}^{\sigma}$, is obtained by tracing the density operator $\hat{\rho}$
of the composite many-body system 
over one of the species, 
while the latter, $\hat{\rho}_1^{\sigma}$, by additionally tracing $\hat{\rho}^{\sigma}$
over all of the particles of the remaining component except one.
In the presence of correlations the resulting reduced density operator
will describe a mixed state. 
The entanglement entropy is caused by intercomponent correlations 
whereas the fragmentation entropy is primarily a signature of intracomponent ones,
though it can be greatly impacted once the intercomponent correlations 
become dominant.
Explicitly, the entanglement and fragmentation entropies are given as:

\begin{equation}
\begin{aligned}
\label{eq:species_entropy}
S_{vN} & = - \tr\left(\hat{\rho}^{\sigma} \ln \hat{\rho}^{\sigma} \right) 
= - \sum_{i=1}^{S} \lambda_i \ln{\lambda_i} \\
&\text{with}  \; \hat{\rho}^{\sigma} 
= \tr_{\bar{\sigma}}(\hat{\rho})
=\sum_{i=1}^{S} \lambda_i \ket{\Psi^{\sigma}_i}\bra{\Psi^{\sigma}_i},
\end{aligned}
\end{equation}

\begin{equation}
\begin{aligned}
	\label{eq:particle_entropy}
S_{vN}^{\sigma} & = 
- \tr\left(\hat{\rho}_1^{\sigma} \ln \hat{\rho}_1^{\sigma} \right)
= - \sum_{i=1}^{s_{\sigma}} n_i^{\sigma} \ln{n_i^{\sigma}}\\
& \text{with}  \ \hat{\rho}_1^{\sigma} 
= \tr_{N_{\sigma}-1}(\hat{\rho}^{\sigma})
=\sum_{i=1}^{s_{\sigma}} n^{\sigma}_i \ket{\Phi^{\sigma}_i}\bra{\Phi^{\sigma}_i}.
\end{aligned}
\end{equation}
In these expressions, $\lambda_i$ and $\ket{\Psi^{\sigma}_i}$ denote the natural populations 
and natural orbitals of the spectrally decomposed $\hat{\rho}^{\sigma}$, while
$ n^{\sigma}_i$ and $\ket{\Phi^{\sigma}_i}$ are the natural populations 
and natural orbitals of the spectrally 
decomposed $\hat{\rho}_1^{\sigma}$~\cite{cao2017unified,alon2021many}. 
Also, $S$ and $s_{\sigma}$ are the number of species orbitals 
and single-particle functions respectively,
$N_{\sigma}$ is the number of $\sigma$ component particles and
$\sigma \neq \bar{{\sigma}}$.

\begin{figure*}[t]
	{\includegraphics[width=0.31\textwidth]{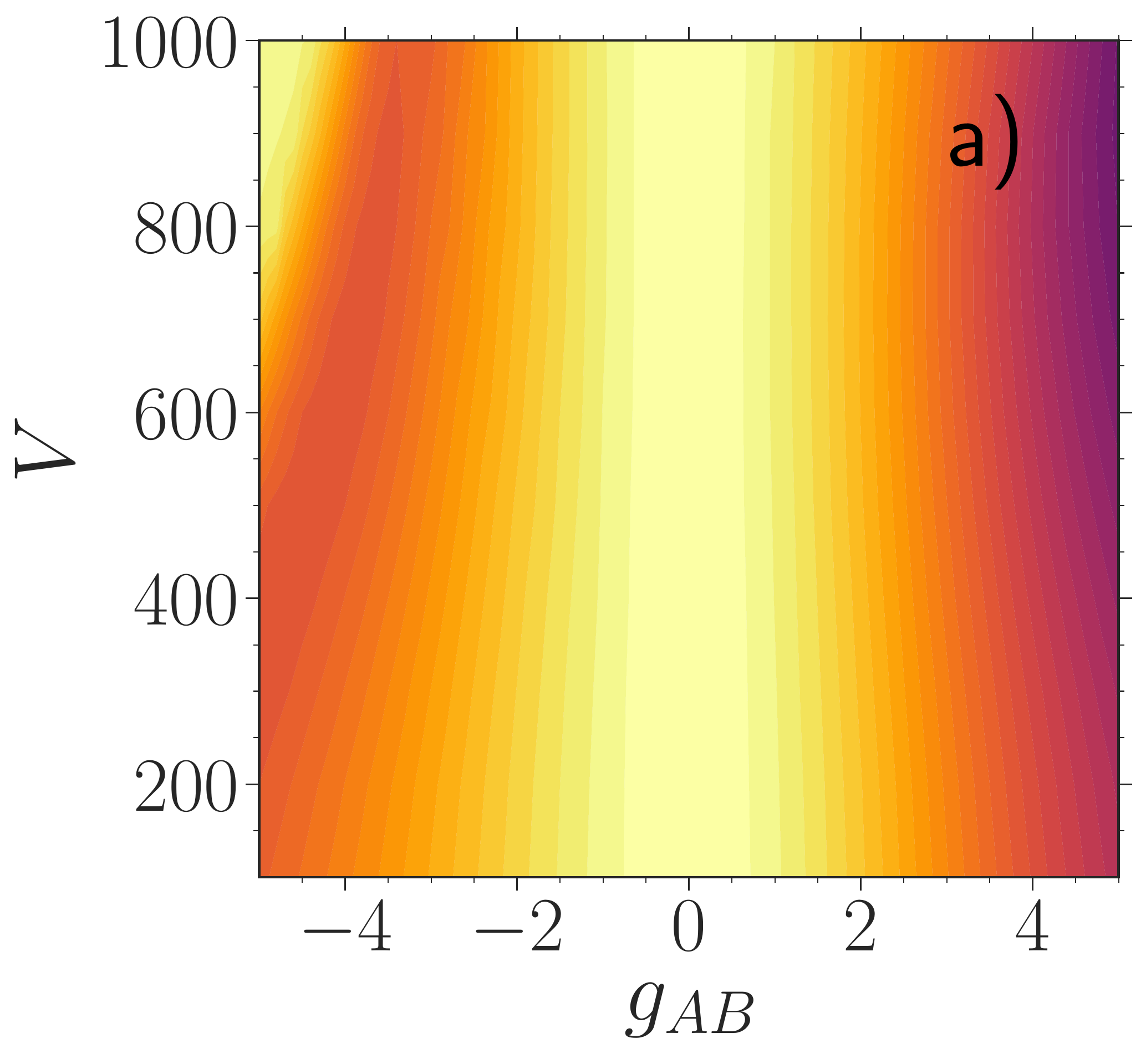}}
	{\includegraphics[width=0.30\linewidth]{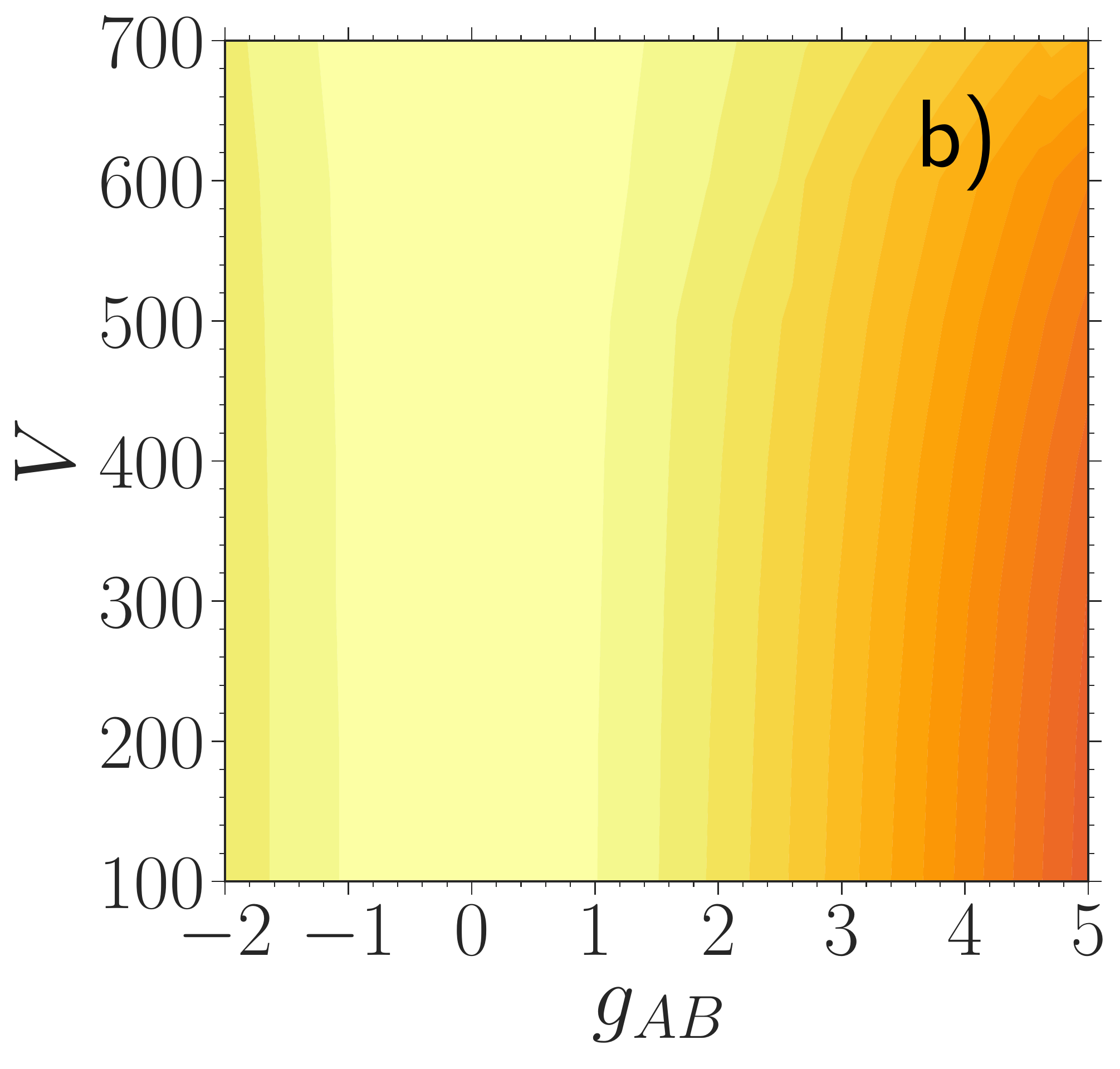}}
	{\includegraphics[width=0.37\linewidth]{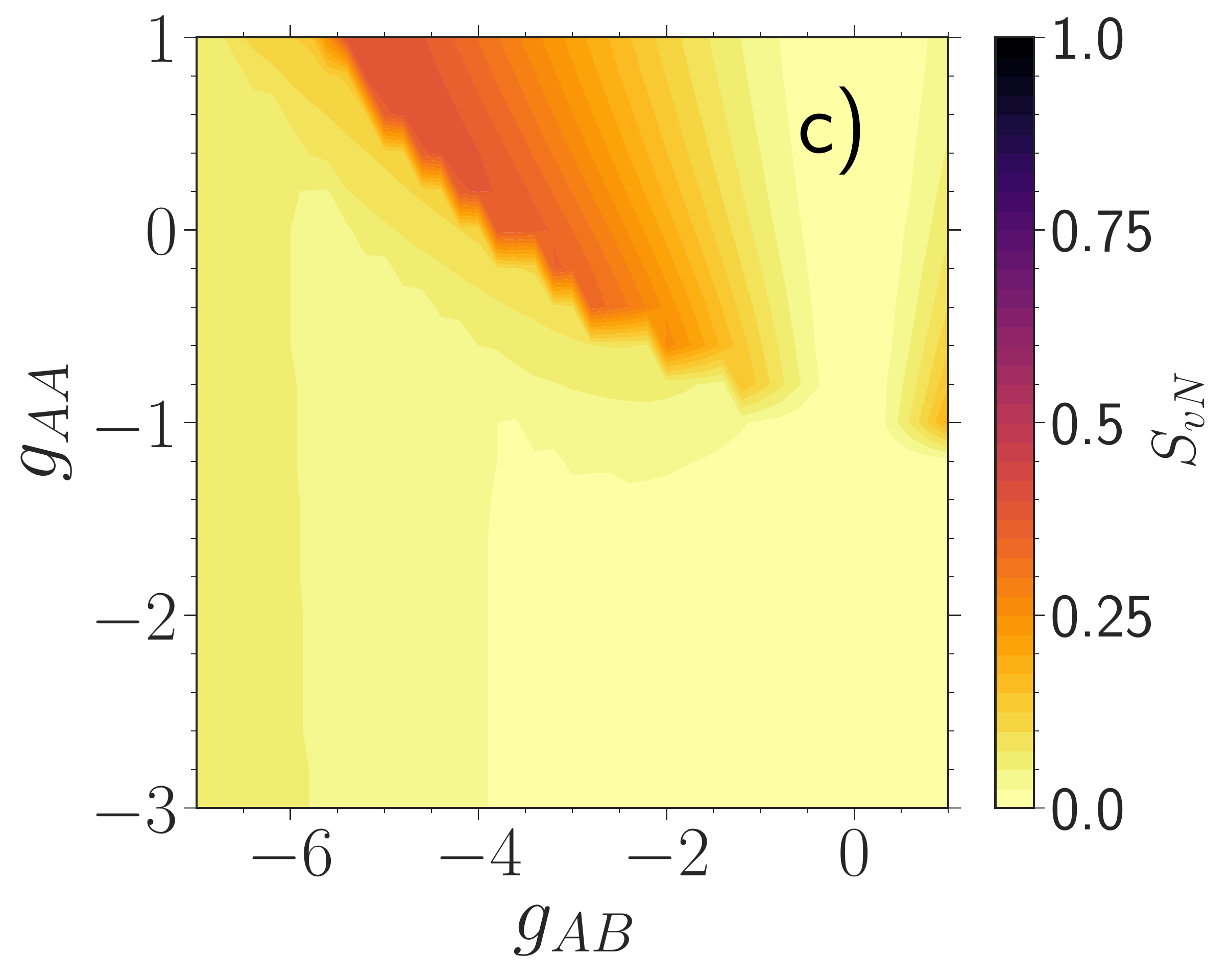}}
%	\begin{subfigure}{.31\textwidth}
%		\centering
%		\includegraphics[width=1\linewidth]{S_vN_gAA05.pdf}
%	\end{subfigure}%
%	\begin{subfigure}{.30\textwidth}
%		\centering
%		\includegraphics[width=1\linewidth]{S_vN_gAA30.pdf}
%	\end{subfigure}
%	\begin{subfigure}{.37\textwidth}
%		\centering
%		\includegraphics[width=1\linewidth]{S_vN_V0500.pdf}
%	\end{subfigure}
%	\begin{subfigure}{.066\textwidth}
%		\centering
%		\includegraphics[width=1\linewidth]{S_vN_cbar.pdf}
%	\end{subfigure}
	\caption{Entanglement entropy $S_{vN}$, 
		see \cref{eq:species_entropy},
		for a) $g_{AA}=0.5$, b) $g_{AA}=3.0$ and c) $V=500$
		with varying majority-impurity coupling $g_{AB}$ 
		and the lattice depth $V$ (a,b) 
		or the interaction strength of the majority atoms $g_{AA}$ (c).
		All quantities are given in box units with characteristic length 
		$R^*=L$ and energy $E^*=\hbar^2/(m L^2)$ 
		with $L$ denoting the extension of the box trap.}
	\label{fig:entanglement_species}
\end{figure*}

\begin{figure*}[t]
	{\includegraphics[width=0.31\textwidth]{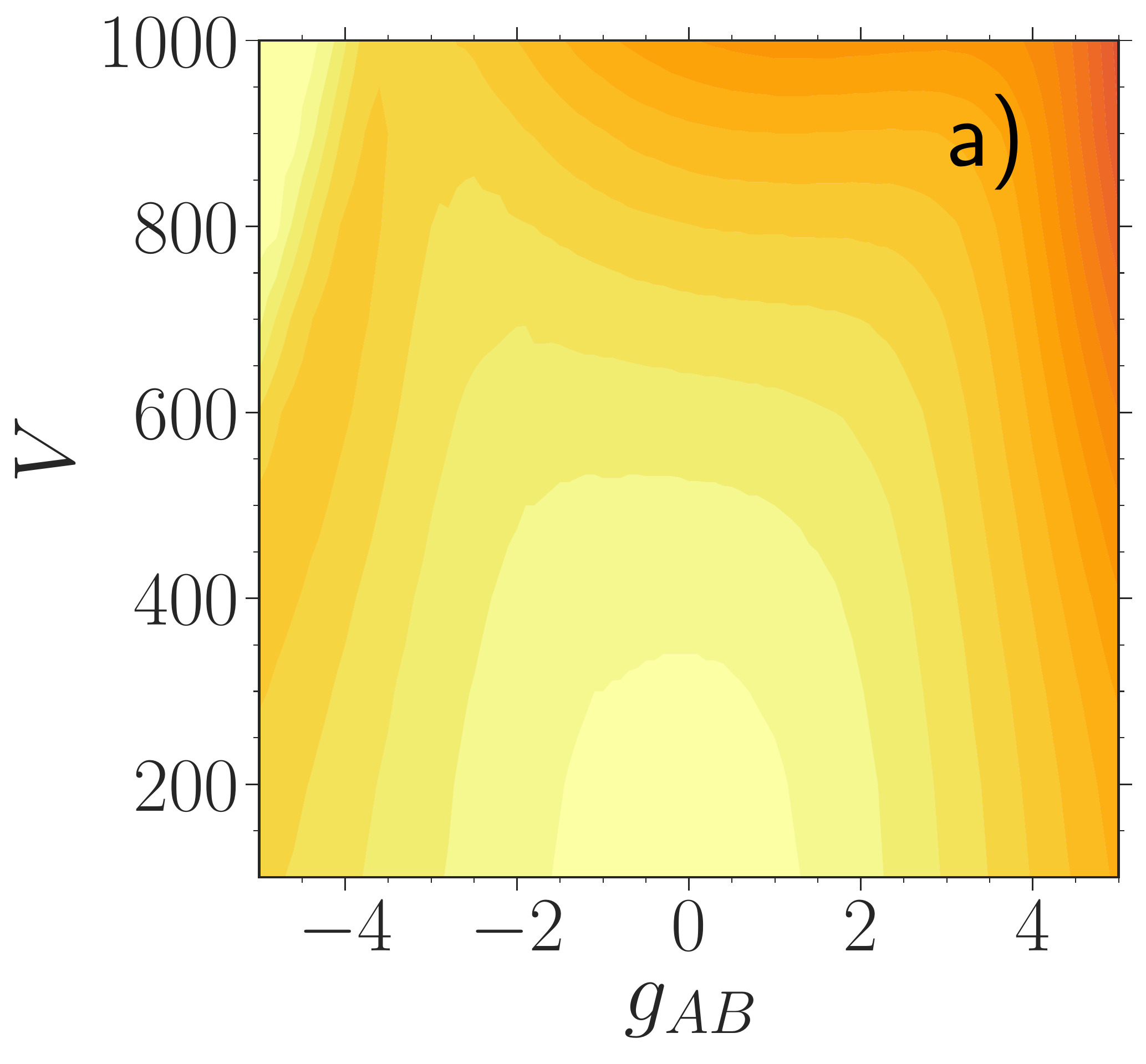}}
	{\includegraphics[width=0.30\linewidth]{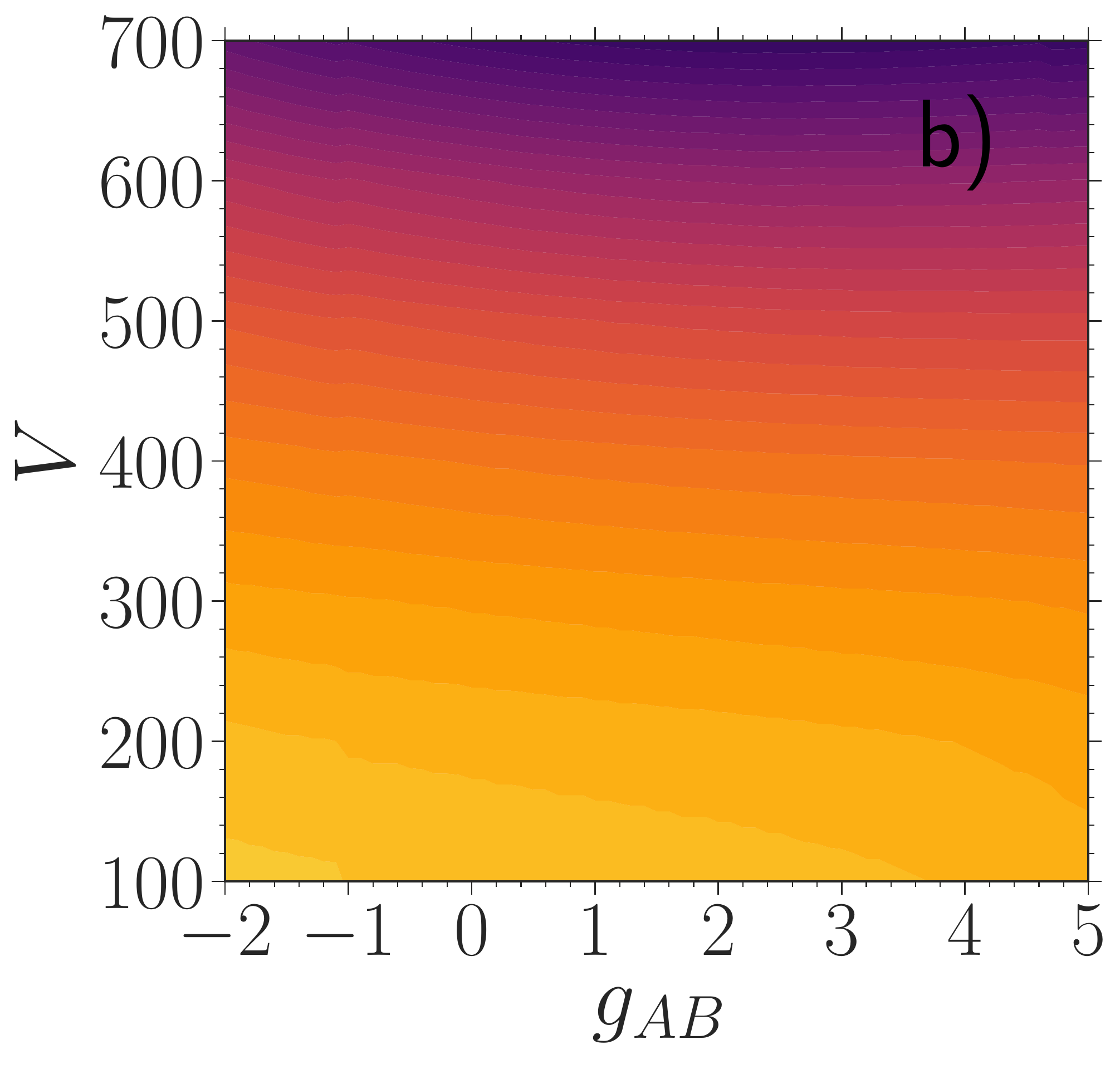}}
	{\includegraphics[width=0.37\linewidth]{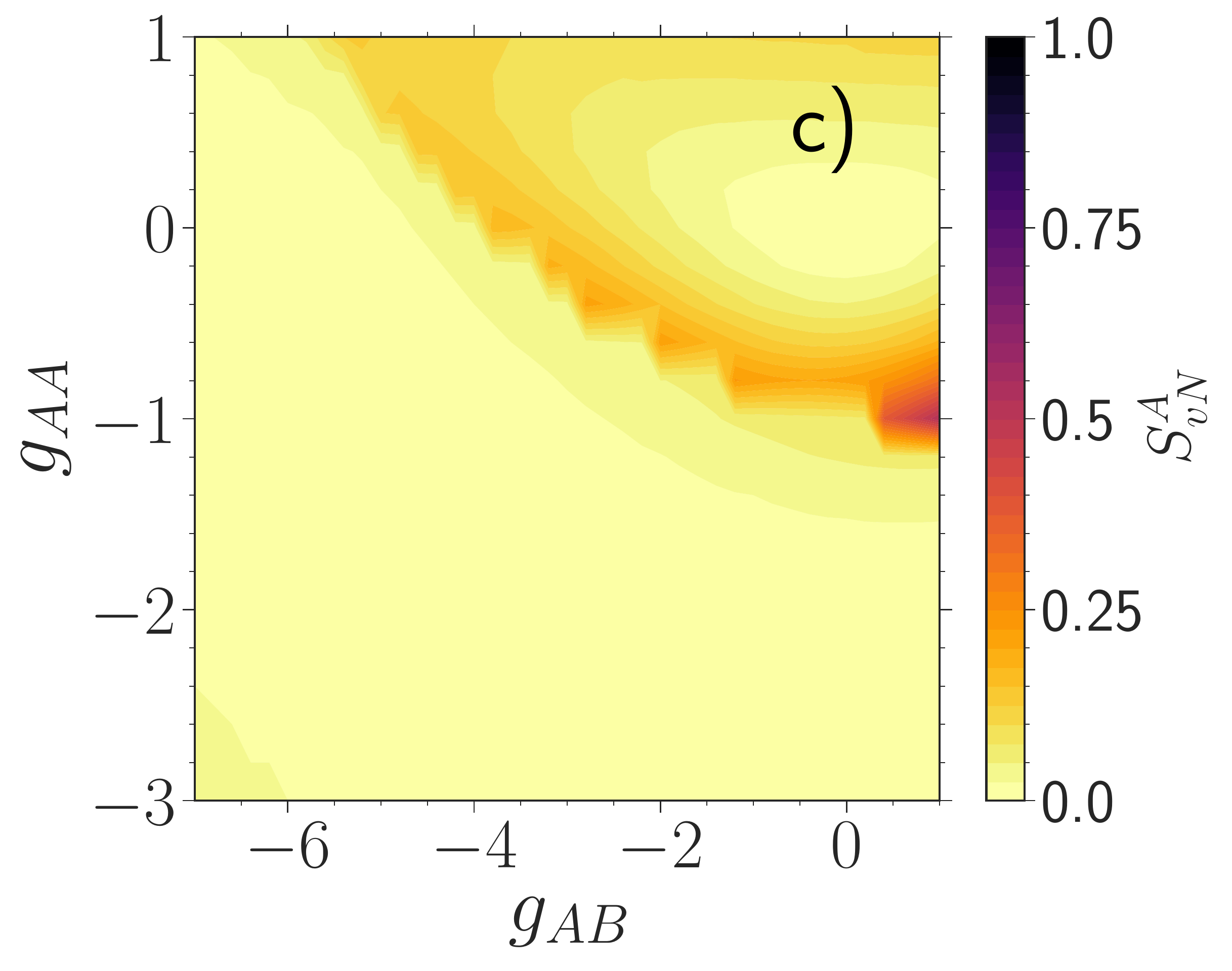}}
%	\begin{subfigure}{.31\textwidth}
%		\centering
%		\includegraphics[width=1\linewidth]{S_vN_A_gAA05.pdf}
%	\end{subfigure}%
%	\begin{subfigure}{.30\textwidth}
%		\centering
%		\includegraphics[width=1\linewidth]{S_vN_A_gAA30.pdf}
%	\end{subfigure}
%	\begin{subfigure}{.37\textwidth}
%		\centering
%		\includegraphics[width=1\linewidth]{S_vN_A_V0500.pdf}
%	\end{subfigure}
%	\begin{subfigure}{.066\textwidth}
%		\centering
%		\includegraphics[width=1\linewidth]{S_vN_A_cbar.pdf}
%	\end{subfigure}
	\caption{Fragmentation entropy $S_{vN}^A$, 
		see \cref{eq:particle_entropy}, 
		for a) $g_{AA}=0.5$, b) $g_{AA}=3.0$ and c) $V=500$
		with respect to the majority-impurity coupling $g_{AB}$ 
		and the lattice depth $V$ (a,b) 
		or the interaction strength of the majority atoms $g_{AA}$ (c).
		All quantities are provided in terms of box units with characteristic length 
		$R^*=L$ and energy $E^*=\hbar^2/(m L^2)$ while $L$ denotes the extension of the box trap.}
	\label{fig:entanglement_fragmentation}
\end{figure*}

In the following, we display the species entanglement $S_{vN}$ from
\cref{eq:species_entropy} [\cref{fig:entanglement_species}] 
and the majority fragmentation  $S_{vN}^A$ from \cref{eq:particle_entropy} 
[\cref{fig:entanglement_fragmentation}]
as a function of the majority-impurity coupling $g_{AB}$ 
and the lattice depth $V$ or the interaction strength of the majority atoms $g_{AA}$.
In case the entanglement entropy $S_{vN}$ is close to zero, 
the corresponding subsystems are to a very good approximation disentangled. 
Thus, making a SMF ansatz in \cref{eq:wfn_ansatz_species_layer}
would greatly facilitate numerical calculations 
while providing quantitatively good results for physical observables.
On the other hand, already moderate values of entanglement 
may have an impact on some physical quantities with measurable differences 
to the SMF approximation.
Regarding the fragmentation entropy of interacting majority atoms $S_{vN}^A$
it is highly non-trivial to predict how their intrinsic mixedness, 
caused by the intra-particle interactions $g_{AA}$,
can be changed by the intercomponent coupling $g_{AB}$.

\subsubsection{Weakly repulsive interacting majority component}

For a weakly interacting majority component with $g_{AA}=0.5$,
the entanglement entropy $S_{vN}$ [\cref{fig:entanglement_species}a]
displays two different behaviors depending 
on the sign of the coupling strength.
For positive $g_{AB}$ it increases gradually 
with increasing coupling strength $g_{AB}$, 
with the build-up being faster for a deeper lattice~\cite{keiler2018correlation}.
This is related to the onset of phase separation 
taking place sooner for a deeper lattice with increasing $g_{AB}$ 
(see also the discussion in \cref{subsec:dmat}).
Turning to negative $g_{AB}$ the entanglement entropy 
first grows gradually with decreasing coupling strength $g_{AB}$,  
but then, for larger $V$ below some threshold value, 
the entanglement reduces to almost zero ($g_{AB}<-4$ and $V>600$).
Apart from the above mentioned pattern
the overall behavior of $S_{vN}$ in \cref{fig:entanglement_species}a is very similar 
to the one observed in the corresponding 
many-body fidelity [\cref{fig:fidelity_full}a].

The fragmentation entropy of the majority component $S_{vN}^A$
[\cref{fig:entanglement_fragmentation}a]
at $g_{AB}=0$ is larger for a deeper lattice.
The reason is that the ratio of the intraspecies interaction energy 
and the single-particle energy of the majority component increases with a larger $V$ or $g_{AA}$. 
In the limit of an infinitely deep lattice 
or an infinitely strong intraspecies repulsion
we expect full fermionization, 
meaning that the one-body density operator becomes a mixed state
with a uniform distribution of natural orbitals
and the fragmentation entropy of the majority component 
reaches the value $\ln(N_A)\approx 1.6$.
However, we observe 
that we are operating far away from that limit, since $\max S_{vN}^{A}<0.4$.

At positive $g_{AB}$, as the entanglement entropy $S_{vN}$ builds up [\cref{fig:entanglement_species}a], 
the fragmentation entropy $S_{vN}^A$ of the majority component at $g_{AB}=0$
is more robust to variations of $g_{AB}$ at deeper lattice depths
compared to shallow lattices [\cref{fig:entanglement_fragmentation}a].
Once the entanglement becomes strong enough 
to overcome intracomponent correlations, 
the fragmentation entropy of the majority atoms starts to increase with a fast rate 
(e.g. $V=1000$, $g_{AB}>4$).
At negative $g_{AB}$, 
if the medium features a small fragmentation entropy at $g_{AB}=0$ ($V<900$),
then $S_{vN}^A$ rises first with decreasing $g_{AB}$, reaches a local maximum 
and finally drops to very small values
at a sufficiently strong coupling strength.
In contrast, if the fragmentation entropy of the decoupled majority component
has already reached a moderate magnitude ($V>900$),
then the initial fragmentation is gradually reduced with decreasing $g_{AB}$, 
until finally both entropies become negligibly small ($g_{AB}<-4$).
Once that happens, the resulting many-body state becomes to a good approximation 
a disentangled composite state with a condensed majority component.

\subsubsection{Moderately repulsive interacting majority component}

The entanglement entropy $S_{vN}$ 
of a moderately interacting majority medium at $g_{AA}=3.0$ 
in \cref{fig:entanglement_species}b 
displays the same qualitative behavior
as the many-body fidelity $F_{mb}$ shown in \cref{fig:fidelity_full}b.
Contrary to $g_{AA}=0.5$ the entanglement
is overall less pronounced and builds up faster at shallow lattice depths instead.
Such a comparatively weak entanglement leaves only a minor imprint 
on the fragmentation of the majority component $S_{vN}^A$, see
\cref{fig:entanglement_fragmentation}b, 
manifested as a weak dependence on the coupling $g_{AB}$.
The fragmentation of the majority species is substantial 
compared to $g_{AA}=0.5$ \cref{fig:entanglement_fragmentation}a at the same lattice depth.
Nevertheless, the fermionization limit is not yet reached, since $\max S_{vN}^{A}\approx0.8$.
The intercomponent correlations
are not strong enough to overcome the intraspecies ones
in accordance with the robustness of the majority component 
observed on the one-body level in \cref{fig:fidelity_A}b.
From this we expect a rather small impact of entanglement on observables, 
which depend solely on the majority particle distribution.

\subsubsection{Attractively interacting majority component}

Finally, we analyze the dependence of the above-described entropy measures
on the intraspecies interaction strength $g_{AA}$
for a moderately deep lattice depth $V=500$ 
[\cref{fig:entanglement_species}c and \cref{fig:entanglement_fragmentation}c].
Since repulsive interactions have been already amply covered, 
we here concentrate on negative $g_{AA}$ and $g_{AB}$.

As it can be readily seen, there is a parameter sector 
at $g_{AB}<0$ and $g_{AA}>-1$
containing high values for the entanglement entropy $S_{vN}$ 
[\cref{fig:entanglement_species}c]. 
This sector displays a similar behavior to $S_{vN}$
in \cref{fig:entanglement_species}a
at negative couplings, namely starting from the decoupled regime, 
the entanglement grows with decreasing $g_{AB}$, 
only to drastically decrease 
below some negative threshold value of $g_{AB}$.
This threshold for $g_{AB}$ lies at lower values 
the higher the intracomponent interaction strength $g_{AA}$ is.
We find that this abrupt decay of $S_{vN}$ coincides with the one observed
in the many-body fidelity $F_{mb}$ [\cref{fig:fidelity_full}c]. 
This suggests that the disappearance of intercomponent correlations
leads to an increased susceptibility of the system to $g_{AB}$ variation.
The other decay region, present in $F_{mb}$ 
at $g_{AA}\approx-1$ and negative $g_{AB}$,
is missing in the entanglement entropy $S_{vN}$.
Form this we infer that it 
can be understood within the SMF picture.
Additionally, there is also another much smaller sector 
characterized by a high entanglement entropy at $g_{AB}>0$ and $g_{AA} \approx -1$.
It is directly related to structural changes observed in $F_{mb}$ and $F_1^A$
at the same values [\cref{fig:fidelity_full}c and \cref{fig:fidelity_A}c], 
which would have been absent in the SMF picture.
Apart from that, below $g_{AA}<-1$ the entanglement entropy among the components 
is either absent or of minor relevance.

Previously, we have mentioned that an isolated majority species,
which interacts repulsively ($g_{AA}>0$), 
features a higher degree of fragmentation
the larger $g_{AA}$ is.
In the case of attractive interactions ($g_{AA}<0$), however, the situation is different.
Namely, starting from $g_{AA}=0$ the fragmentation entropy 
tends first to increase with decreasing $g_{AA}$, 
but then decreases up to the point of describing approximately
a condensed state again [see \cref{fig:entanglement_fragmentation}c at $g_{AB}=0$].
Regarding the impact of the intercomponent coupling $g_{AB}$ on $S_{vN}^A$ 
we observe overall very similar patterns as for the entanglement entropy $S_{vN}$
[\cref{fig:entanglement_species}c].
Regions where both entropic measures $S_{vN}$ and $S_{vN}^A$ 
are of small magnitude remind of the corresponding sectors 
in \cref{fig:entanglement_species}a and \cref{fig:entanglement_fragmentation}a
at $V>800$ and $g_{AB}<-4$.

\newpage

% % % % % % % % % % % % % % % % % % % % % % % % % % % % % % % % %
\subsection{Single- and two-particle density distributions}
\label{subsec:dmat}

The measures of fidelity and entropy
discussed in the previous sections
are very useful in identifying parameter regions 
being substantially impacted and/or highly correlated
indicating regimes of high interest for further investigation.
However, they do not provide insights into the actually undergoing processes.
To get a better understanding we ask for the impact on
measurable quantities such 
as the one-body and two-body density distribution functions,
which can be accessed by fluorescence imaging 
with a quantum gas microscope \cite{Bakr2010,Sherson2010,Omran2015,Hohmann2017,pyzh2019quantum}.

In the following,
$\rho_1^{\sigma} (z)$ describes the probability density
to find a single particle of species $\sigma$ at position $z$, while
$\rho_2^{\sigma \bar{\sigma}} (z_1, z_2)$ denotes the probability density
to simultaneously measure one particle of species $\sigma$ 
at position $z_1$
and another one of the same or different species $\bar{\sigma}$ 
at position $z_2$. 
The expectation value of any local observable
depending on up to two degrees of freedom 
can be evaluated as an overlap integral 
with the appropriate probability density.
Since many local observables often depend only on the distance
between the particles, i.e. $O(z_1, z_2)=O(z_1-z_2)$, 
we replace $\rho_2^{\sigma \bar{\sigma}} (z_1, z_2)$ by
the probability density $\rho_r^{\sigma \bar{\sigma}} (r)$ 
to measure two particles belonging to the same or different species
at a relative distance $r$ 
independent of their individual positions.
To this end we perform a coordinate transformation 
$R=(z_1+z_2)/2$ and $r=z_1-z_2$ giving the following identity:
\begin{flalign}
\int \rho_2^{\sigma \bar{\sigma}} (z_1, z_2) \, dz_1 \, dz_2 
& = \int \rho_2^{\sigma \bar{\sigma}} (r, R) \, dr \, dR.
\end{flalign}
Then we define:
\begin{flalign}
\label{eq:dmatAB_rel}
\rho_r^{\sigma \bar{\sigma}} (r) 
& = \int \rho_2^{\sigma \bar{\sigma}} (r, R) \, dR.
\end{flalign}
Our first goal here is to investigate how the above mentioned quantities are affected 
in parameter sectors displaying strong susceptibility to structural changes
identified in \cref{subsec:fidelity} and, in particular, 
whether the density distributions are capable to capture 
the undergoing changes in the many-body state.

Our second goal is to extract the impact of the entanglement. 
To this end we compare the above density distributions
obtained from the variational ML-X calculations 
to the ones where the SMF ansatz is assumed.
The latter will be distinguished by a tilde sign placed 
on top of the corresponding quantities.
In the following, we shall evince
that a large entanglement entropy identified in \cref{subsec:entropy}
has indeed a notable impact, but not always on all
of the above mentioned density distributions.
Thus, it may enhance or impede the effects coming from the induced SMF potential,
such as phase separation and localization, 
or affect the bunching properties of the majority component.

\subsubsection{Weakly repulsive interacting majority component}

\begin{figure*}[t]
	\centering
	\includegraphics[width=1.0\textwidth]{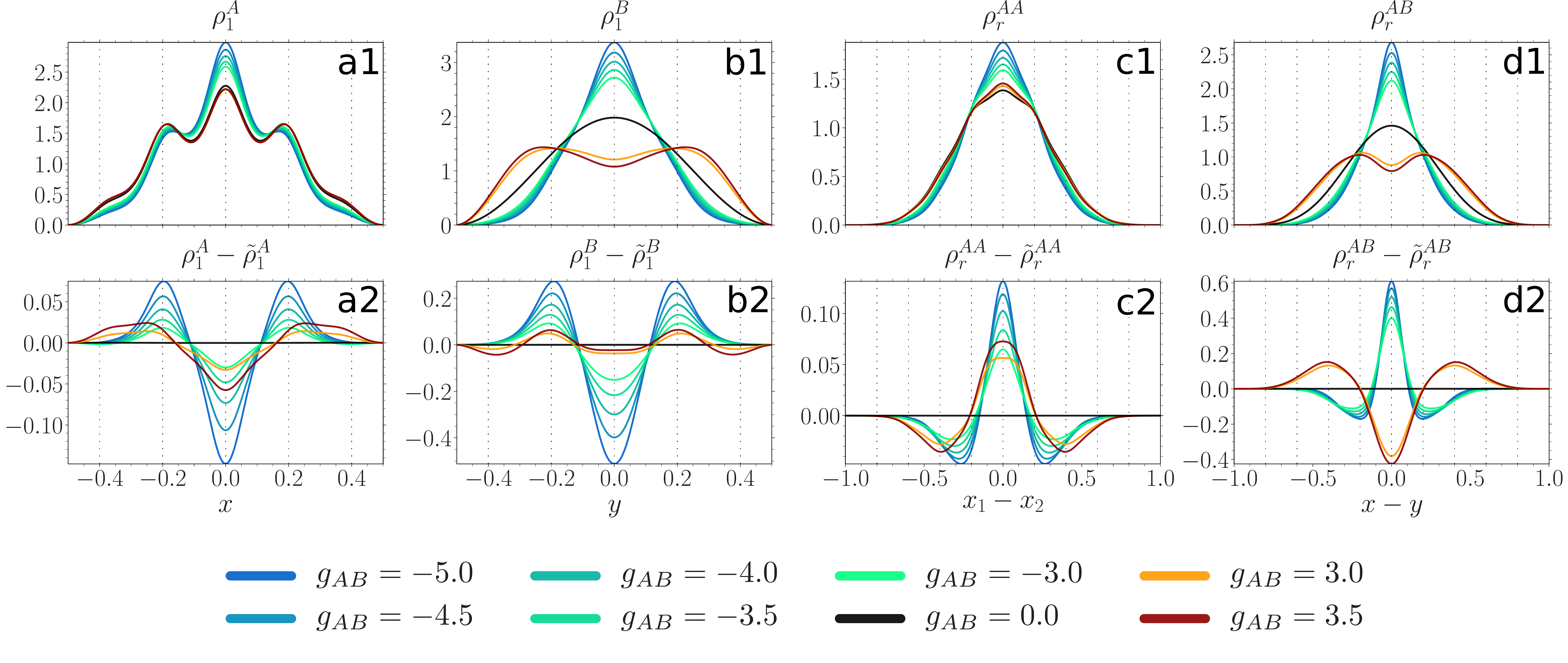}
	\caption{
		Upper panels: one-body probability densities
		$\rho_1^{A}(x)$, $\rho_1^{B}(y)$ [\cref{eq:rho1}] 
		and distance probability distributions
		$\rho_r^{AA}(x_1-x_2)$, $\rho_r^{AB}(x-y)$ [\cref{eq:dmatAB_rel}]
		at $g_{AA}=0.5$, $V=100$ and for various values of $g_{AB}$ (see legend).
		Lower panels: difference between probability densities obtained from the
		variational ML-X simulations and the SMF ansatz, 
		the latter distinguished by a tilde sign.
		All quantities are given in box units with characteristic length 
		$R^*=L$ and energy $E^*=\hbar^2/(m L^2)$ 
		with $L$ being the extension of the box trap.
	}
	\label{fig:gpops_gAA05_V0100.eps}
\end{figure*}

For a shallow lattice ($V=100$) we observe
in \cref{fig:gpops_gAA05_V0100.eps} that
the majority component (panel a1) at $g_{AB}=0$
occupies mainly the {\it central} site (at $z=0$)
and the two {\it intermediate} ones (at $z=\pm0.2$), 
while $\rho_r^{AA}$ (panel c1) features an almost Gaussian shape 
due to weak intraspecies correlations.
At moderate positive couplings ($g_{AB}>3$) 
both quantities are only slightly affected
in accordance with the robustness of $F_1^A$ 
in this interaction regime [\cref{fig:fidelity_A}a].
At moderate negative couplings ($g_{AB}<-3$) both $\rho_1^A$ and $\rho_r^{AA}$ shrink
with decreasing $g_{AB}$ indicating an increased bunching tendency 
of the majority atoms towards the central lattice site.
The impact of entanglement here is moderate.
It leads to an increased probability
for the majority component to occupy the two intermediate sites,
while disfavoring the central site (panel a2). 
Thus, it acts as an inhibitor 
of localization at negative $g_{AB}$ and
counteracts changes induced by the SMF potential at positive $g_{AB}$.
Furthermore, entanglement favors bunching of the majority particles
independent of the sign of the coupling (panel c2).

The decoupled impurity particle (panel b1) occupies 
the ground state of the box potential. 
At moderate positive couplings it develops two humps and
forms a shell around the majority component density, 
a signature of phase separation~\cite{keiler2020doping,mistakidis2019quench} further confirmed by the appearance of two humps
in $\rho_r^{AB}$ (panel d1).
At negative couplings $\rho_1^B$ and $\rho_r^{AB}$
shrink with decreasing $g_{AB}$ accumulating around the trap center.
The entanglement favors the process of phase separation 
at positive couplings and bunching between 
the two species at negative couplings (panel d2), 
while slowing down the shrinking of $\rho_1^B$ 
at negative coupling (panel b2).
We also remark that upon reaching 
a certain threshold value of $g_{AB}>4$,
the SMF solution experiences 
breaking of parity symmetry,
causing substantial differences 
to the many-body symmetry-preserving solution (not shown).

\begin{figure*}[t]
	\centering
	\includegraphics[width=1.0\textwidth]{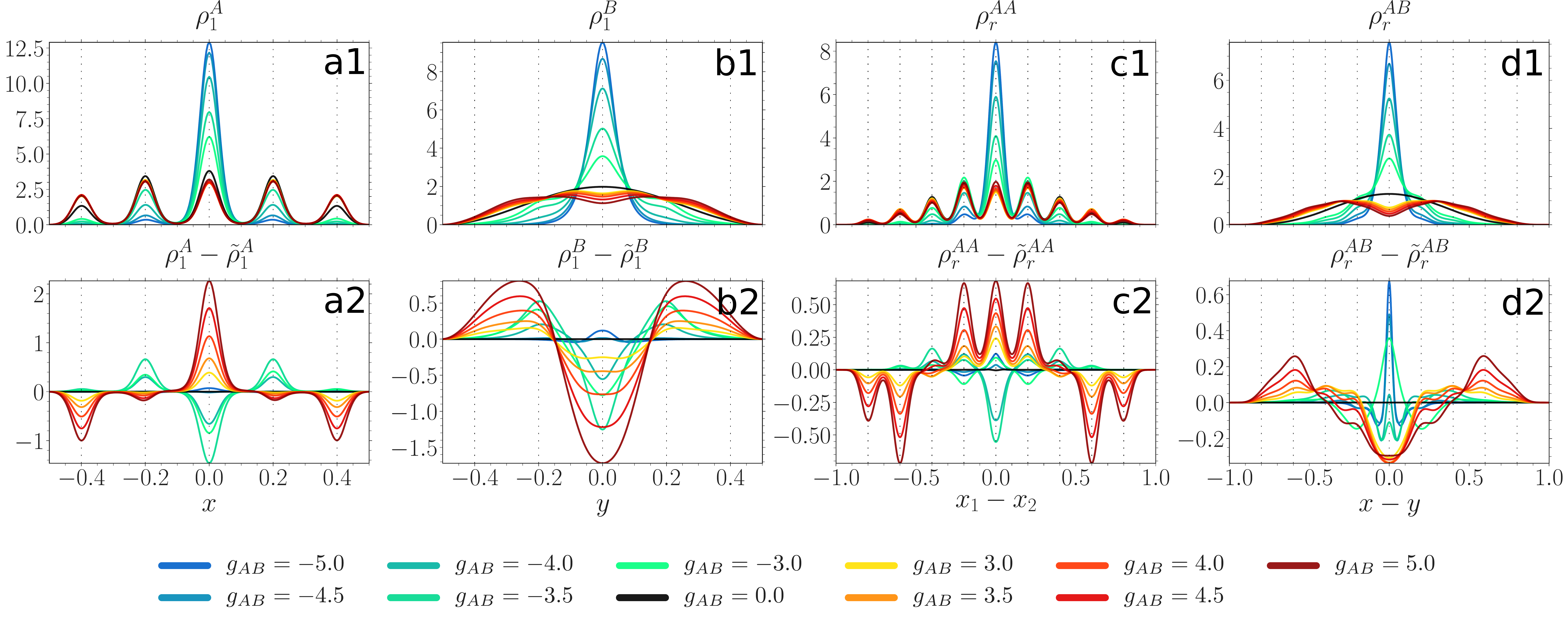}
	\caption{
		Upper panels: one-body probability densities
		$\rho_1^{A}(x)$, $\rho_1^{B}(y)$ [\cref{eq:rho1}] 
		and distance probability distributions
		$\rho_r^{AA}(x_1-x_2)$, $\rho_r^{AB}(x-y)$ [\cref{eq:dmatAB_rel}]
		at $g_{AA}=0.5$, $V=1000$ and for various values of $g_{AB}$ (see legend).		
		Lower panels: difference between probability densities obtained from
		many-body ML-X calculations and SMF ansatz, 
		the latter distinguished by a tilde sign.
		All quantities are given in box units with characteristic length 
		$R^*=L$ and energy $E^*=\hbar^2/(m L^2)$ 
		with $L$ denoting the extension of the box trap.
	}
	\label{fig:gpops_gAA05_V01000.eps}
\end{figure*}

For a deep lattice ($V=1000$) 
in \cref{fig:gpops_gAA05_V01000.eps}
the majority component (panel a1) at $g_{AB}=0$
displays an almost uniform distribution over all the lattice sites, 
while $\rho_r^{AA}$ (panel c1) features a multi-hump structure 
due to stronger intraspecies correlations 
[cf.\ \cref{fig:entanglement_fragmentation}a)].
At moderate positive couplings ($g_{AB}>3$) 
the width of $\rho_1^{A}$ and $\rho_r^{AA}$ is only slightly increased,
again in accordance with the robustness of $F_1^A$ [\cref{fig:fidelity_A}a]. 
Thus, the majority component, 
experiencing the presence of a repelling impurity atom, 
shows a slight enhancement of the already present delocalization over the lattice.
At moderate negative couplings ($g_{AB}<-3$) both $\rho_1^A$ and $\rho_r^{AA}$ shrink
with decreasing $g_{AB}$ to the extent 
where all atoms occupy predominantly only the central site ($g_{AB}<-4$).
Such a large difference to the non-interacting ground state 
is in accordance with the observations made 
in $F_1^A$ [\cref{fig:fidelity_A}a].

The impact of entanglement 
is structurally different compared to a shallow lattice (panels a2 and c2).
At positive couplings, 
the entanglement greatly increases the probability 
for the majority atoms to be found at the central site,
while decreasing the probability at {\it outer} sites ($z=\pm0.4$) 
and being indifferent to the intermediate sites (panel a2).
Additionally, it favors the bunching of the majority particles 
at the same or neighboring sites 
and disfavors them being more than two sites apart (panel c2).
At negative couplings, it acts in a similar way 
as in the case of shallow lattices, 
except that for a sufficiently strong coupling strength 
($g_{AB}<-4$), where both entropy measures are of small magnitude 
[see \cref{fig:entanglement_species,fig:entanglement_fragmentation}a],
the SMF ansatz is in good accordance 
with the many-body solution. 

The impurity particle (panel b1) at positive couplings ($g_{AB}>3$)
first develops two humps,
but then as the coupling increases,
the relative distance between those peaks grows, 
while the humps themselves become flatter.
There is a strong signature of an onset of a four-peak structure at $g_{AB}=5$.
This is in accordance
with the increasing relative distance between the species (panel d1)
and the fact that the majority atoms are distributed uniformly 
over all the lattice sites in contrast to $g_{AA}=0.5$,
where the majority component was occupying 
mainly the central and the intermediate sites.
At negative couplings ($g_{AB}<-3$) $\rho_1^B$ and $\rho_r^{AB}$
shrink with decreasing $g_{AB}$.

The entanglement favors the process where the impurity atom moves
from the box center to its boundaries
independently of the sign of the coupling (panel b2).
At $g_{AB}<-4.0$ it plays only a minor role, the same as for the majority component.
Regarding $\rho_r^{AB}$, at positive couplings the entanglement favors 
the process of phase separation 
by pushing the impurity particle more than two sites apart 
from a majority atom (panel d2).
At negative couplings it enhances the bunching between 
the two species, even when the entanglement entropy 
is very small (e.g.\ at $g_{AB}=-5.0$).

\subsubsection{Moderately repulsive interacting majority component}

\begin{figure*}[t]
	\centering
	\includegraphics[width=1.0\textwidth]{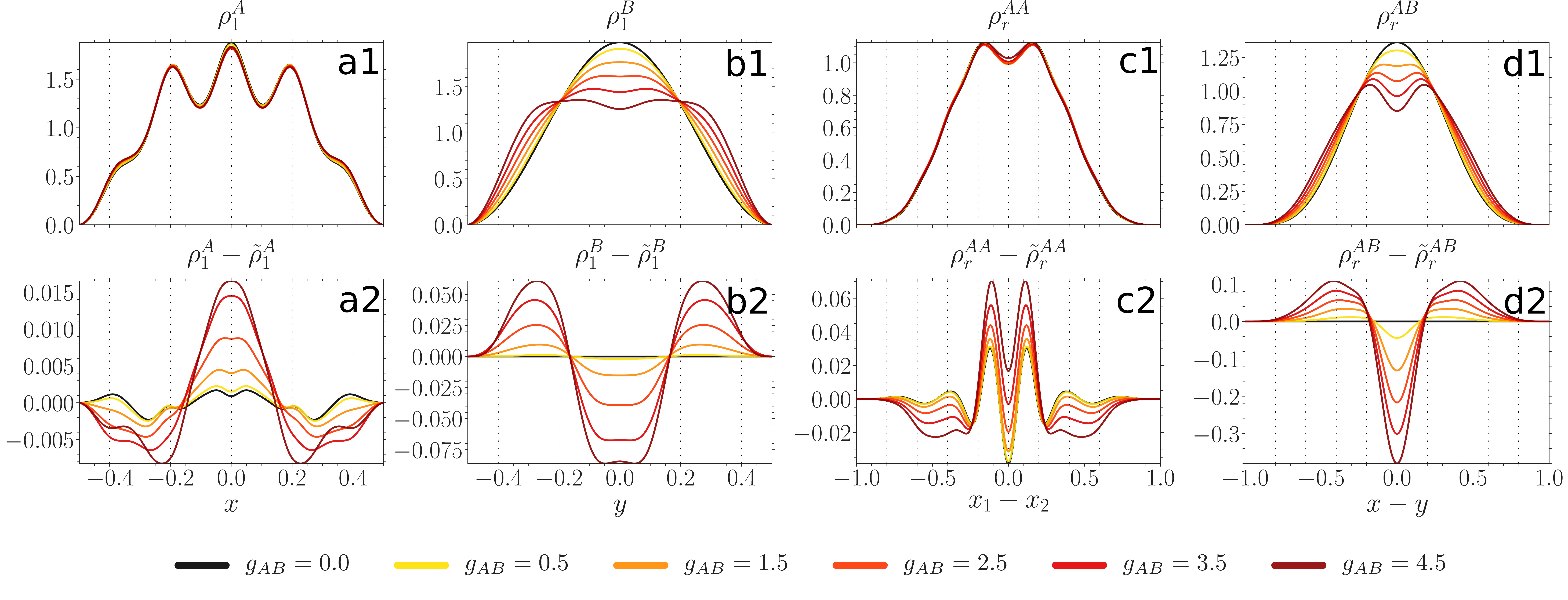}
	\caption{
		Upper panels: one-body probability densities
		$\rho_1^{A}(x)$, $\rho_1^{B}(y)$ [\cref{eq:rho1}] 
		and distance probability distributions
		$\rho_r^{AA}(x_1-x_2)$, $\rho_r^{AB}(x-y)$ [\cref{eq:dmatAB_rel}]
		at $g_{AA}=3.0$, $V=100$ and for different values of $g_{AB}$ (see legend).		
		Lower panels: difference between probability densities obtained from
		the many-body ML-X calculations and SMF ansatz, 
		the latter distinguished by a tilde sign.
		All quantities are provide in terms of box units with characteristic length 
		$R^*=L$ and energy $E^*=\hbar^2/(m L^2)$ while $L$ is the extension of the box trap.
	}
	\label{fig:gpops_gAA30_V0100.eps}
\end{figure*}

Considering our findings regarding fidelity and entropy measures 
we investigate here only shallow lattices at positive couplings 
(\cref{fig:gpops_gAA30_V0100.eps}), 
where the structural changes caused by the coupling and 
the entanglement entropy $S_{vN}$ may have a sizable impact on
density distributions.
The decoupled density of the majority component (panel a1) 
has three pronounced humps 
at the {\it central} ($z=0$) and {\it intermediate} sites ($z=\pm0.2$). 
The profile is overall more spread compared to a weakly interacting majority 
[cf.\ \cref{fig:gpops_gAA05_V0100.eps} panel a1].
Indeed, it is most beneficial for two particles 
to occupy neighboring sites (see the two humps in panel c1).
The majority component gets only a weak feedback from 
the presence of a repulsive impurity atom, even 
at coupling strengths comparable to $g_{AA}$
in accordance with the robustness of $F_1^A$ in \cref{fig:fidelity_A}b.
The role of the entanglement is also rather weak,
though qualitatively different to $g_{AA}=0.5$ in \cref{fig:gpops_gAA05_V0100.eps}.
Thus, it increases the probability for the majority particle to be found
at the region enclosed between the two intermediate sites, 
while decreasing the probability to be detected outside of that region (panel a2).
Furthermore, it favors particle distances 
of a half lattice constant ($a_l=0.2 R^*$) (panel c2).

The impurity particle (panel b1) experiences phase separation 
similar to \cref{fig:gpops_gAA05_V0100.eps} (panel b1),
i.e.,\ upon increasing $g_{AB}$
it develops two humps with a minimum at the trap center.
Then, those humps separate and flatten, until finally they would form
a four-hump structure with three local minima located 
at the position of the three peaks in the majority component density 
(compare to panel a1).
The separation between the species 
is again clearly manifested as two humps in $\rho_r^{AB}$
with favored distance of a lattice constant ($a_l=0.2 R^*$) (panel d1).
The entanglement
affects the impurity atom in quite an opposite way
when compared to the majority component (panel b2),
i.e., it decreases the probability for the impurity atom to be found
at the region enclosed between the two intermediate sites, 
while increasing the probability to lie outside of that region.
Additionally, similar to the behavior at weaker $g_{AA}$ 
[cf.\ \cref{fig:gpops_gAA05_V0100.eps} panel d2],
the entanglement accelerates the phase separation process (panel d2).

\subsubsection{Attractively interacting majority component}

\begin{figure*}[t]
	\centering
	\includegraphics[width=1.0\textwidth]{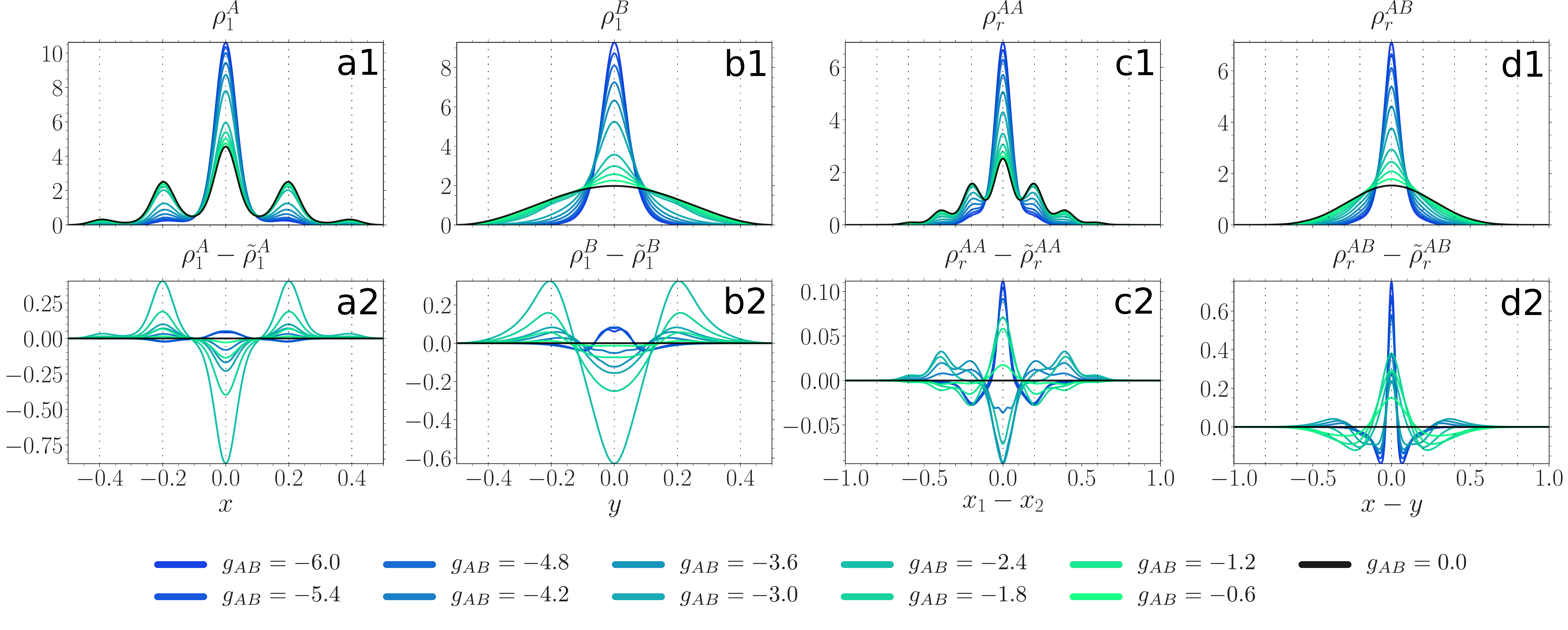}
	\caption{
		Upper panels: one-body probability densities
		$\rho_1^{A}(x)$, $\rho_1^{B}(y)$ [\cref{eq:rho1}] 
		and distance probability distributions
		$\rho_r^{AA}(x_1-x_2)$, $\rho_r^{AB}(x-y)$ [\cref{eq:dmatAB_rel}]
		at $g_{AA}=-0.4$, $V=500$ and for various values of $g_{AB}$ (see legend).	
		Lower panels: difference between probability densities obtained from
		the variational ML-X simulations and SMF ansatz, 
		the latter distinguished by a tilde sign.
		All quantities are expressed in box units of characteristic length 
		$R^*=L$ and energy $E^*=\hbar^2/(m L^2)$ 
		while $L$ being the extension of the box trap.
	}
	\label{fig:gpops_gAA-04_V0500.eps}
\end{figure*}

Finally, we concentrate on negative 
intraspecies interactions $g_{AA}$, namely
a weak negative $g_{AA}=-0.4$ 
at negative $g_{AB}$ [\cref{fig:gpops_gAA-04_V0500.eps}],
contained in the parameter sector 
with substantial entanglement entropy
[\cref{fig:entanglement_species}c].

In \cref{fig:gpops_gAA-04_V0500.eps} 
a decoupled majority atom where $g_{AB}=0$
is localized at the {\it central} ($z=0$) and {\it intermediate} ($z=\pm0.2$) wells (panel a1). 
Even though the majority atoms are attracted to each other,
the probability to be one or even two wells apart is still sizable (panel c1).
With decreasing $g_{AB}$ both $\rho_1^A$ and $\rho_r^{AA}$
shrink to a Gaussian.
The impact of entanglement is quite different 
compared to the previously considered cases.
Thus, at $g_{AB}>-4.8$ the entanglement slows down
the process of $\rho_1^A$ localization at the central well (panel a2).
The strongest impact is reached around $g_{AB} \approx -2.4$,
where the entanglement entropy is largest 
for the given value of intracomponent interaction $g_{AA}=-0.4$ 
[\cref{fig:entanglement_species}c].
Below $g_{AB}<-4.8$, as the entanglement entropy suddenly drops, so does
the difference to the SMF ansatz.
The intercomponent correlations favor clustering of the majority atoms
at $-2.4<g_{AB}<0$ and $g_{AB}<-4.2$, 
whereas at $-4.2<g_{AB}<-2.4$, where the entanglement entropy is largest,
they inhibit the clustering (panel c2).

The impurity density $\rho_1^B$ 
shows a similar behavior as the majority component
density (panel b1), also in terms of the role of the entanglement (panel b2).
The width of $\rho_r^{AB}$ shrinks with decreasing $g_{AB}$ (panel d1), 
while the entanglement enhances the bunching between 
the two species (panel d2).
\begin{figure*}[t]
	\includegraphics[width=\textwidth]{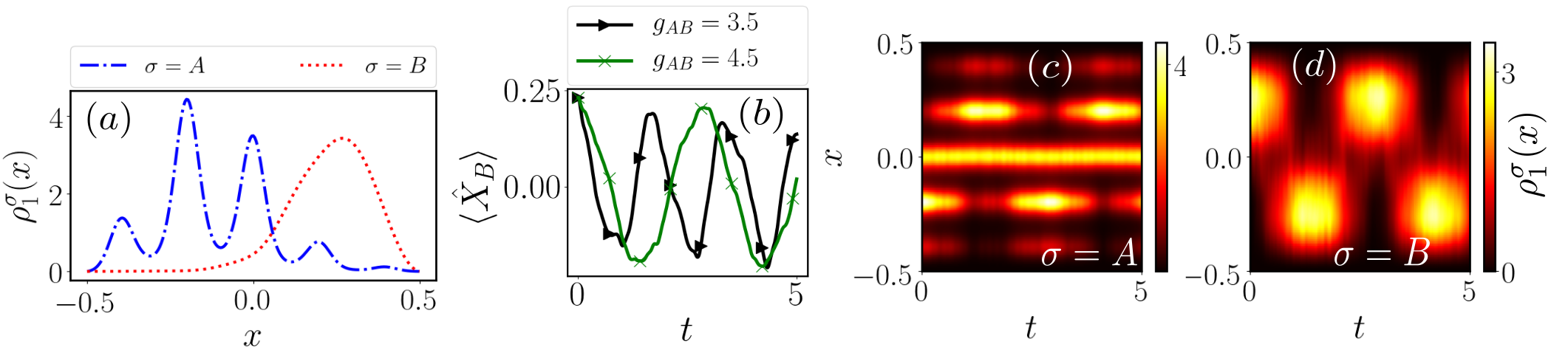}
	\caption{(a) One-body density $\rho^{\sigma}_1(x)$ of the initial state configuration for $V=500$, $g_{AA}=0.5$ and $g_{AB}=6.0$ at $t=0$. Temporal evolution of (b) the averaged position of the impurity $\langle\hat{X}_B\rangle$ [see eq.~(\ref{eq:com})], (c) the one-body density of the majority species and (d) the one-body density of the impurity upon quenching the interspecies interaction strength to $g_{AB}=4.5$. \label{fig:densities}}
\end{figure*}
\section{Quench induced tunneling dynamics}
\label{sec:dyn}
Having analyzed in detail the ground state properties of our system, we subsequently study the dynamical response of a single impurity coupled to a lattice trapped species upon quenching the interspecies interaction strength $g_{AB}$. To this end we prepare the system in its ground state for $V=500$, $g_{AB}=6.0$ and $g_{AA}=0.5$, leading to the formation of a two-fold degeneracy in the ground state and the two species phase separate~\cite{keiler2020doping}. In this sense, the ground state one-body density is given by a superposition state of two parity-symmetry broken configurations, where the density of the first one is depicted in Fig. \ref{fig:densities}a and the second one corresponds to its parity-symmetric (with respect to $x = 0$) counterpart.
It is possible to remove this degeneracy in order to select any of the states in the respective degenerate manifold. Technically, this is done by applying a small asymmetry, e.g. a tilt, to the lattice potential, thereby breaking the parity symmetry and energetically favoring one of the above-mentioned states~\cite{theel2020entanglement}.

\begin{figure*}[t]
	\includegraphics[width=0.8\textwidth]{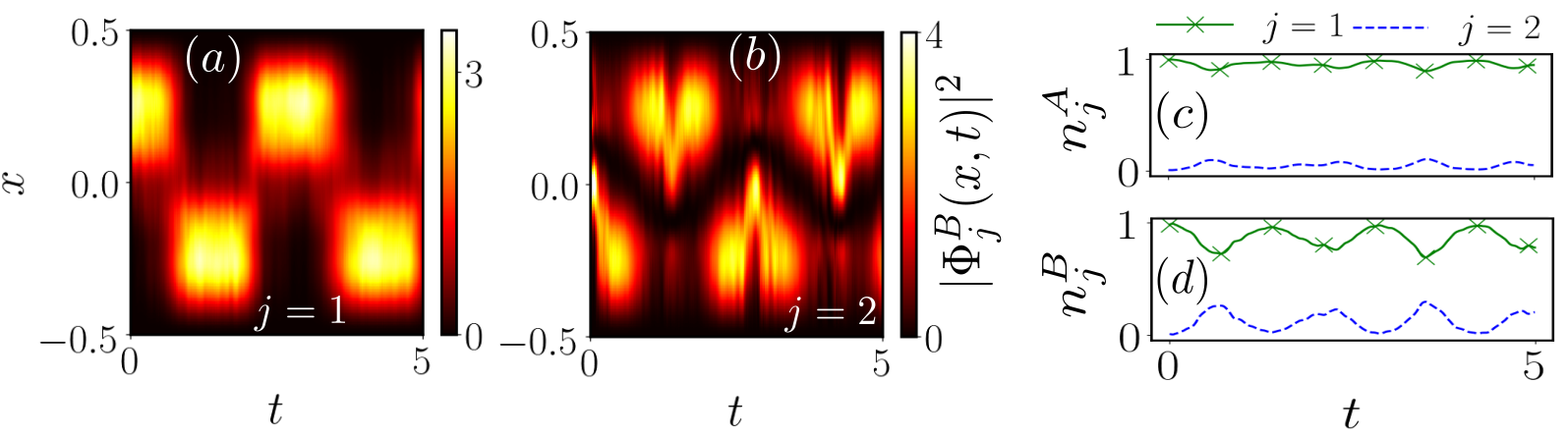}
	\caption{Temporal evolution of the density of (a) the first and (b) the second natural orbital $\Phi^{B}_j(x,t)$ [see eq.~(\ref{eq:particle_entropy})] of the impurity and (c), (d) the natural populations $n^{\sigma}_j$ of both subsystems upon quenching the interspecies interaction strength $g_{AB}$ of the ground state in Fig. \ref{fig:densities}a to $g_{AB}=4.5$. \label{fig:natorb}}
\end{figure*}

To trigger the dynamics starting from the initial state configuration illustrated in Fig. \ref{fig:densities}a we quench the interspecies interaction strength to a smaller value. 
As a representative example of the emergent tunneling dynamics of each species we present the temporal evolution of the corresponding one-body densities in Fig. \ref{fig:densities}c,d following a quench to $g_{AB}=4.5$, while keeping fixed $V=500$ and $g_{AA}=0.5$. In this case the impurity performs an oscillatory motion which is reminiscent of the tunneling of a particle in a double-well. This can be attributed to the lifting of the degeneracy for smaller interspecies interaction strengths. For a post-quench value of $g_{AB}=4.5$ the initially prepared state has a substantial overlap with the post-quench ground state and the first excited state such that in the course of the dynamics the system will oscillate between those two. This is similar to a single particle in a double-well which is prepared as a superposition of the first doublet and undergoes a tunneling between the sites.
Correspondingly, the majority species will undergo a collective tunneling in the lattice geometry~\cite{theel2020entanglement,keiler2019interaction}. Thus, the probability distribution of a single majority species particle will oscillate between the initial distribution [Fig. \ref{fig:densities}a] and its parity-symmetric counterpart. Due to the repulsive nature of the interspecies coupling the two species move in opposite directions such that they end up in phase-separated configurations after half a period. 
Note that the oscillation period, being the energy gap between the two energetically lowest eigenstates of the post-quench Hamiltonian (not shown here), depends on the post-quench $g_{AB}$. This can be easily verified by monitoring the temporal evolution of the averaged position of the impurity~\cite{catani2012quantum} which is defined as 
\begin{equation}
	\langle\hat{X}_B\rangle=\int_{-L/2}^{L/2}dx\rho_{1}^{B}(x) x.
	\label{eq:com}
\end{equation}
For various post-quench $g_{AB}$ we find that the impurity will occupy its parity-symmetric counterpart, reflected in the decrease of $\langle\hat{X}_B\rangle$ towards negative values, while the oscillation decreases with smaller $g_{AB}$ [Fig. \ref{fig:densities}b].  
In order to gain insight into beyond mean-field effects we investigate the natural populations $n^{\sigma}_j$ [see eq.~(\ref{eq:particle_entropy})] which indicate the degree of fragmentation of the subsystem~\cite{lode2020colloquium,mistakidis2018correlation}.
For simplicity here we present the populations of the first two dominantly populated natural orbitals while using six orbitals in the actual calculations. 
The initial depletion of both subsystems is rather small, i.e. $n^{A}_1\approx0.996$ and $n^{B}_1\approx0.99$, such that any decrease of these populations upon quenching $g_{AB}$ is due to dynamical many-body effects.
We find that for both subsystems dominantly two natural orbitals contribute during the dynamics [Fig. \ref{fig:natorb}c,d], while the ones of the medium are less impacted by the quench. For the natural populations of the impurity signatures of an oscillation can be observed, where $n^B_1$ initially decreases and revives back towards $n^{B}_1\approx0.99$, while $n^B_2$ initially increases and afterwards drops back to nearly zero. In order to attribute the occupation of the additional natural orbital to physical processes, we analyze the spatial distribution of the natural orbitals $\Phi^{B}_j(x,t)$ [see eq.~(\ref{eq:particle_entropy})] themselves focusing on the impurity [Fig. \ref{fig:natorb}a,b].
In Fig. \ref{fig:natorb}a we observe that the first natural orbital corresponds to the oscillatory behavior of the one-body density  of the impurity, but lacking the smooth transition between the phase-separated configurations [see Fig. \ref{fig:densities}d]. The first natural orbital dominates during the dynamics and we can interpret its behavior as corresponding to the presence of the phase-separated density configurations. Consequently, the second natural orbital [Fig. \ref{fig:natorb}b], resembling the mirror image of the first one, contributes to deviations from this solution. Due to its structure we can deduce that it is responsible for initiating the transport of the impurity, thereby allowing for the counterflow of the two species.
Note that the presence of more than one natural orbital during the dynamics is a clear signature that mean-field theory would not provide an accurate description of the system dynamics. Hence, the fact that $\ket{\Phi^B_2}$ is occupied is a manifestation of many-body effects, influencing the motion of both species.

%Quenching the interspecies interaction strength we are able to induce a dynamical process which for the impurity is reminiscent of the tunneling of a single particle in a double well potential.
%This can be attributed to the lifting of the quasi-degeneracy for the corresponding post-quench Hamiltonian. Due to the repulsive interspecies interaction also the majority species will undergo a tunneling in the lattice geometry such that the two species move in opposite directions, ending up in phase-separated configurations after half a period.
%We identify the presence of two dominant natural orbitals for the impurity species during the dynamics, where the first one corresponds to the dynamical behavior of the respective one-body density while the second one resembles the mirror image of the first one. The presence of an additional natural orbital emphasizes the many-body character of the dynamics, thereby influencing the motion of the impurity.

\section{Summary and Outlook}
\label{sec:summary}

In this work we analyze the static and dynamical properties of a few-body
particle-imbalanced bosonic mixture at zero temperature.
Importantly, the components are exposed to different one-dimensional 
external traps where the majority species is subject to a finite lattice potential 
while the single impurity is trapped in a box of the same extension as the lattice.
We study the response of the composite system 
upon the variation of majority-impurity coupling $g_{AB}$ and
majority component internal parameters being either the lattice depth $V$
or the majority-majority interaction strength $g_{AA}$.

To quantify the response of static properties
we employ the fidelity between two density operators describing
ground states at zero and a finite intercomponent interaction $g_{AB}$.
We contrast the response at the many-body to the single-particle level.
We observe that the composite system is quite robust to the variation of the intercomponent interaction
at strongly repulsive $g_{AA}$, while being fragile 
at strongly attractive $g_{AB}$ and deep lattices $V$ as well as 
when $g_{AA}$ is weakly attractive and $g_{AB}$ is strongly attractive.
Upon comparison to the fidelities between the corresponding reduced one-body density operators
of each component, 
we not only observe that each species is affected to a much smaller degree,
but they also respond differently.
Thus, for the impurity atom the deviation from the box ground state 
increases smoothly with increasing absolute value of $g_{AB}$, 
while the reduced density of the majority component remains very robust to $g_{AB}$ variations
except for the above mentioned parameter regions where the many-body fidelity exhibits
significant structural changes in the ground state.

Next, we have been performing a further classification of our system based 
on entropy measures.
Namely, we quantify the amount of entanglement 
and intraspecies correlations deposited in the binary mixture 
by evaluating the von-Neumann entropy of the respective subsystems.
Interestingly, we find that our composite system is only weakly entangled 
for parameter regions which undergo substantial structural changes.
Additionally, we observe
that while the entanglement entropy continuously grows with increasing repulsive $g_{AB}$,
it does not behave the same for attractive $g_{AB}$, where
it reaches a local maximum at a finite value of $g_{AB}<0$.
Another peculiar observation is that the fragmentation entropy of the majority component
undergoes a strong variation
for parameter regions, where the fidelity measure does not show any evidence
of majority particles being affected by the intercomponent interaction.
Even though the mixed character of the reduced density of the medium 
suffers from substantial changes,
it remains un-observable on the single-particle level.

To visualize our observations stemming from the fidelity measure
we show the one-body density distributions of each component along
with the probability distributions for two particles of the same or different species to
be measured at a relative distance from each other.
These quantities are usually accessible in state-of-the-art ultracold atom experiments
and determine the expectation values of local one- and two-body observables.
Indeed, strong deviations appearing in the fidelity at the single-particle level
are also clearly visible in the corresponding one-body density.
At positive couplings
we observe an interspecies phase separation where the impurity is pushed to the box edges, while leaving the majority component intact.
At negative couplings both components tend to increase their localization
at the central well.

To further quantify our conclusions stemming from the entanglement measure
we rely on the difference between the above probability distributions and 
the corresponding ones when assuming a disentangled state in our calculations.
Again, we find strong deviations for parameters displaying high entanglement entropy values.
Thus, at positive couplings the entanglement favors the process of phase separation,
while at negative couplings it generally, but not always, 
counteracts the localization of both species.

Quenching the interspecies interaction strength we are able to induce a dynamical process which for the impurity is reminiscent of the tunneling of a single particle in a double well potential.
This can be attributed to the lifting of the degeneracy for the corresponding post-quench Hamiltonian as well as the substantial overlap of the initial state configuration with the post-quench ground state and the first excited state. Due to the repulsive interspecies interaction also the majority species will undergo a tunneling in the lattice geometry such that the two species move in opposite directions, ending up in phase-separated configurations after half a period.
We identify the presence of two dominant natural orbitals for the impurity species during the dynamics, where the first one corresponds to phase-separated configurations in the respective one-body density, while the second one resembles the mirror image of the first one. The presence of an additional natural orbital emphasizes the many-body character of the dynamics, thereby influencing the motion of the impurity. 

There are various promising research directions that are worth pursuing in the future. 
Indeed, the generalization of our findings for an increasing particle number 
in the medium or larger lattice potentials as well as the role of the lattice filling factor 
is desirable. 
Also, a more elaborated analysis on the possibly emerging impurity-medium bound states 
or the engineering of droplet-like configurations in such settings at strong intercomponent attractions 
would be important. 
Furthermore, it would be intriguing to study the persistence and possible alterations 
of the identified spatial configurations in the presence of finite temperature 
which will impact the coherence of the lattice bosons~\cite{lingua2017thermometry,mahmud2011finite}. Another perspective is to investigate the relevant radiofrequency spectrum~\cite{mistakidis2020radiofrequency,mistakidis2020pump} 
in order to capture the emergent polaron properties including their lifetime,
residue and effective mass especially in the attractive interaction regimes of bound state formation.

\begin{acknowledgments}
	M.\ P.\ and K.\ K.\ gratefully  acknowledge a scholarship  
	of the Studienstiftung des deutschen Volkes.
	S.\ I.\ M.\ gratefully acknowledges financial support in the framework of the Lenz-Ising Award of the Department of Physics of the University of Hamburg.
\end{acknowledgments}

\bibliography{references}
\bibliographystyle{apsrev4-2}

\end{document}